\documentclass{elsart}

\usepackage{psfig}

\begin{document}

\runauthor{MZB}
\begin{frontmatter}

\title{ Stochastic Renormalization Group in Percolation:
I. Fluctuations and Crossover  }

\author[MIT]{Martin Z. Bazant}
\address[MIT]{Department of Mathematics, Massachusetts Institute of
Technology, Cambridge, MA 02139}


\date{June 10, 2002}


\begin{abstract}
A generalization of the Renormalization Group, which describes
order-parameter fluctuations in finite systems, is developed in the
specific context of percolation.  This ``Stochastic Renormalization
Group'' (SRG) expresses statistical self-similarity through a
non-stationary branching process. The SRG provides a theoretical basis
for analytical or numerical approximations, both at and away from
criticality, whenever the correlation length is much larger than the
lattice spacing (regardless of the system size). For example, the SRG
predicts order-parameter distributions and finite-size scaling
functions for the complete crossover between phases. For percolation,
the simplest SRG describes structural quantities conditional on
spanning, such as the total cluster mass or the minimum chemical
distance between two boundaries. In these cases, the Central Limit
Theorem (for independent random variables) holds at the stable,
off-critical fixed points, while a ``Fractal Central Limit Theorem''
(describing long-range correlations) holds at the unstable, critical
fixed point. This first part of a series of articles explains these
basic concepts and a general theory of crossover. Subsequent parts
will focus on limit theorems and comparisons of small-cell SRG
approximations with simulation results.
\end{abstract}

\end{frontmatter}


\begin{center} 
Dedicated to H. Eugene Stanley on the occasion of his
sixtieth birthday\footnote{ This paper was originally
prepared for the proceedings of the conference, ``Horizons in Complex
Systems,'' held in Messina, Sicily in December 2001 to commemorate
Prof. Stanley's birthday. }.
\end{center}

\vskip 24pt

\section{Introduction}

The theory of critical phenomena is mostly concerned with phase
transitions in the thermodynamic limit of infinite systems. In this
limit, an order parameter has unambiguous values in each phase and is
singular at a precise critical point of the control variable
(temperature, concentration, etc.)~\cite{stanley,ma,goldenfeld,kadanoff}.
In finite systems, however, there are two basic modifications to the
thermodynamic picture: (i) The critical point becomes blurred since
all statistical averages are analytic functions of the control
variable, and (ii) the order parameter fluctuates around its mean
value.  Accordingly, the order parameter in a finite system is
described by a {\it probability distribution}, which depends
non-trivially on the control variable and the system size.

As arguably the simplest critical phenomenon,
percolation~\cite{bunde,hughes,stauffer} provides a natural setting
within which study order-parameter fluctuations in finite
systems. Moreover, in recent years there has been growing practical
interest in the distribution functions of structural
quantities, such as the minimum or maximum chemical distance between
two
terminals~\cite{havlin85,havlin87a,havlin87,neumann88,hovi95,hovi97,dok98,dok99,andrade00,paul02},
the masses of spanning clusters~\cite{neumann88,hovi95,hovi97,sen99},
and the mass of the largest
cluster~\cite{duxbury87,bazant00,zeifman00,sen01}.  These random
quantities not only act as order parameters for the phase transition,
but also play a crucial role in both
dynamical~\cite{king99,lee99,andrade00} and
mechanical~\cite{duxbury87,zeifman00} properties of heterogeneous
materials~\cite{torquato}.

There is no existing theory to predict order-parameter statistics for
arbitrary values of concentration and system size, but considerable
progress has been made in understanding the limiting behavior in each
thermodynamic phase. Classical limit theorems for independent random
variables hold in the normal phases, where correlations are negligible
(by definition). In the supercritical phase, the masses of large
fragments of the infinite cluster~\cite{newman80,newman81a,newman83}
and other structural quantities~\cite{cox81,cox84} satisfy the Central
Limit Theorem, i.e. the distribution is Gaussian, and the mean and
variance both grow linearly with the system size. In the subcritical
phase, the mass of the largest cluster is governed by the theory of
extremes of independent random variables~\cite{duxbury87,bazant00}. In
this case, the distribution has the Fisher-Tippett (or Gumbel) form
with non-universal corrections due to discrete-lattice effects, while
the mean grows logarithmically with system size (with a bounded
variance)~\cite{bazant00}.

Computer simulations of percolation have shown that various order
parameters in the critical phase have different universal
distributions with stretched-exponential tails (when scaled to have
unit
mean)~\cite{havlin85,havlin87a,neumann88,sen99,dok98,dok99,andrade00,sen01}.
These signatures of what we call the ``Fractal Central Limit Theorem''
~\cite{bb02} can be predicted by analyzing {\it stationary} branching
processes~\cite{feller,harris,athreya}, as was recently pointed out by
Hovi and Aharony~\cite{hovi97}.  Aside from an early analysis of
mean-field percolation on the Cayley tree~\cite{havlin87}, these
authors made the first (and, to the author's knowledge, only)
systematic use of branching processes to approximate critical limiting
distributions, although they focused mainly on the tails of
spanning-cluster mass distributions~\cite{hovi97}.

In this first part of a series of articles, we present the
``Stochastic Renormalization Group'' (SRG), a theoretical formalism
based on {\it non-stationary} branching processes which describes the
{\it crossover} of order-parameter distributions between different
thermodynamic phases.  The SRG expresses the statistical
self-similarity of finite systems in which the correlation length,
$\xi$, and linear system size, $L$, are both much larger than the
lattice spacing, $a$. The critical ($\xi \gg L$) and off-critical
($\xi \ll L$) fixed points of the SRG correspond to various limit
theorems for the order parameter. In the latter case, small
correlation lengths ($\xi \approx a$) are also allowed.  Although the
basic idea of the SRG applies to critical phenomena and random
fractals in general, we develop it here on the specific context of
percolation.

The article is organized as follows.  In section~\ref{sec:RG}, we set
the stage by reviewing classical renormalization-group (RG) concepts
in percolation. In section~\ref{sec:SRG}, we present the SRG for any
statistical quantity conditional on spanning the system, such as the
total mass of spanning clusters or the minimum chemical distance. In
section~\ref{sec:cross}, we show how the SRG generally describes
crossover. In sections~\ref{sec:CLT} and \ref{sec:FCLT}, we briefly
explain why these the off-critical and critical fixed points are
associated with the CLT and the FCLT, respectively. In
section~\ref{sec:disc}, we summarize the SRG ``flow'' in the space
probability distributions for percolation quantities dependent upon
spanning, and we conclude by discussing the general relevance of the
SRG for fractals and phase transitions in finite systems. Subsequent
parts will analyze the fixed points in detail, including more
comparisons with simulation data.  The subtle case of the largest
cluster (which may not span), the most natural order parameter in
finite
systems~\cite{duxbury87,bazant00,stauffer80,family80,margolina84},
will also be addressed.

\section{ RG Concepts in Percolation }
\label{sec:RG}

\subsection{ Self-Similarity at the Critical Point }

The basic idea behind all RG methods for critical phenomena is the
{\it statistical self-similarity of large systems near a critical
point}~\cite{ma,goldenfeld,kadanoff}. This essentially follows from
dimensional analysis~\cite{barenblatt87}: In the limit $a/L\rightarrow
0$ and $a/\xi\rightarrow 0$, there are only two (out of originally
three) relevant length scales, $L$ and $\xi$, so all systems with the
same dimensionless ratio, $L/\xi$, must have approximately the same
statistical properties (upon a suitable normalization). The
self-similarity becomes exact in the thermodynamic limit, $L
\rightarrow \infty$, as the critical point is approached, $\xi
\rightarrow \infty$, with the scaling variable, $L/\xi$, held
constant.

The notion of self-similarity is exploited by the RG to calculate
nontrivial properties of the critical point, which cannot be obtained
by simple dimensional analysis.  For example, in the case of
percolation, according to the arguments above, a large system of $N =
mn$ sites must have approximately the same connectivity statistics as
a coarse-grained system of $n \gg 1$ large ``cells'' (or ''blocks'')
containing $m \gg 1$ sites each, provided that an effective occupation
probability for cells, $p_m = R_m(p)$, is defined so as to preserve
the dimensionless ratio, $L/\xi$:
\begin{equation}
\frac{(mn)^{1/d}}{\xi(p)} = \frac{n^{1/d}}{\xi(p_m)}. \label{eq:xisim}
\end{equation}
where $d$ is the system dimension ($N = L^d$).  This is analogous
to Kadanoff's original block renormalization of the coupling
constant for magnetic spins~\cite{kadanoff}.

If the exact dependence of the correlation length on $p$ were known,
then the cell occupation probability would be given by
\begin{equation}
R_m(p) = \xi^{-1}\left(\frac{\xi(p)}{m^{1/d}}\right).
\end{equation}
The trick is to now reverse the logic: From a suitable definition of
the cell occupation probability, an estimate of the correlation
length, $\xi^{(n)}(p)$, can be obtained by successively
``renormalizing" the occupation probability,
\begin{equation}
p_{mn} = R_n(p_m), \label{eq:pRG}
\end{equation}
starting from $p_1=p$. By construction, an appropriate choice of
$R_n(p)$ yields a valid asymptotic approximation in the large-cell
limit, $\lim_{n \rightarrow\infty} \xi^{(n)}(p) = \xi(p)$, although it
is not trivial to make this choice for a given problem such that the
convergence is fast enough to allow the use of small cells.

\subsection{ Asymptotic Approximations }

\begin{figure}
\begin{center}
\mbox{
\psfig{file=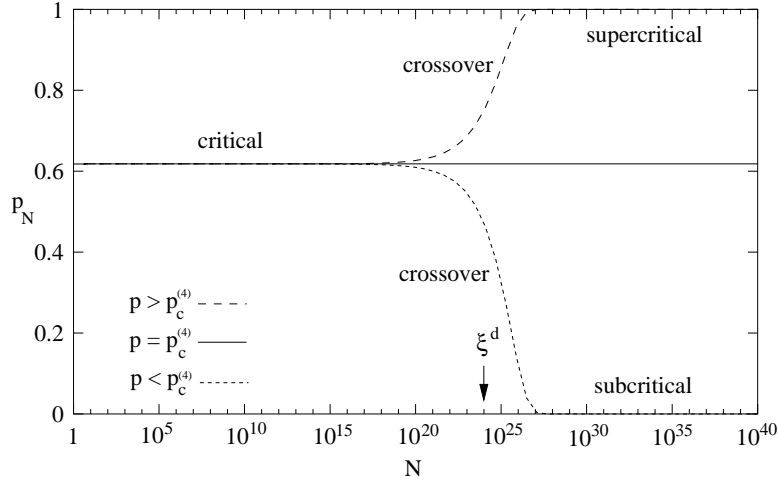,height=2.5in}
}
\begin{minipage}[h]{5.5in}
\caption{ Typical trajectories of the renormalized cell occupation
probability $p_N$ given by Eq.~(\protect\ref{eq:pRG}) for the
four-site ($N = 4^j$) RG scheme described below in
Eq.~(\protect\ref{eq:R4}) starting at the critical fixed point, $p_1 =
p_c^{(4)} = (\sqrt{5}-1)/2$ (solid line), a nearby subcritical value,
$p_1^\prime = p_c^{(4)} - 10^{-8}$ (dotted line), and a nearby
supercritical value, $p_1^{\prime\prime} = p_c^{(4)} + 10^{-8}$
(dashed line). An associated estimate of the correlation length,
$\xi^{(4)}(p_1^\prime)$, in the subcritical case is labeled as the
system size $N = \xi^d$ where the renormalized occupation probability
$p_N$ drops below 0.5.  This is very close to a correlation length
estimate, $\xi^{(4)}(p_1^{\prime\prime})$, for the supercritical case
where $p_N$ rises above 0.7. \label{fig:pn} }
\end{minipage}
\end{center}
\end{figure}

Estimates of correlation length and $p_c$ itself are obtained from the
RG, Eq.~(\ref{eq:pRG}), as follows. As shown in Fig.~\ref{fig:pn}, for
any suitable choice of $R_n(p)$, the recursion (\ref{eq:pRG})
converges to one of three possible fixed points in the thermodynamic
limit, $j \rightarrow\infty$:
\begin{equation}
\lim_{j\rightarrow \infty} p_{n^j} = \left\{ \begin{array}{ll}
0 & \mbox{ for }  0 \leq p < p_c^{(n)}\\
p_c^{(n)} & \mbox{ for }  p = p_c^{(n)}\\
1 & \mbox{ for }  p_c^{(n)} < p \leq 1\\
\end{array} \right.
\end{equation}
the stable subcritical fixed point, $p=0$; the stable supercritical
fixed point, $p=1$; or the unstable critical fixed point, $p =
p_c^{(n)}$, which is an $n$-cell estimate of the true infinite-system
value of $p_c$. As long as the RG transformation, is chosen
to satisfy
\begin{equation}
\lim_{n\rightarrow \infty} R_n(p) = \left\{ \begin{array}{ll} 0 &
\mbox{ for } 0 \leq p < p_c \\ c & \mbox{ for } p = p_c \\ 1 & \mbox{
for } p_c < p \leq 1
\end{array} \right.   \label{eq:Rlimit}
\end{equation}
for some constant $0 \leq c \leq 1$, then the RG approximation
converges in the large-cell limit, $\lim_{n \rightarrow\infty}
p_c^{(n)} = p_c$. 
As shown in Fig.~\ref{fig:pn}, for $p \neq p_c^{(n)}$, an estimate of
the correlation length, $\xi^{(n)}(p)$, can be defined as the
(interpolated) linear system size, $n^{j/d}$, where the cell
occupation probability, $p_{n^j}$, first passes within a given
tolerance of a stable fixed point as $j \rightarrow\infty$.

The RG introduces a useful artificial length scale, the linear cell
size, $b$ ($n = b^d$). In order for the lattice-spacing $a$ to not
affect the RG calculation of critical behavior, we must have $b\gg a$,
but the underlying assumption of self-similarity also requires $b \ll
\xi(p)$, which is guaranteed near the critical point, where $\xi(p_c)
= \infty$. General considerations of dimensional analysis imply that
such a divergence must have the form of a power law, $\xi(p) \sim
(p-p_c)^{-\nu}$ (e.g. for $p \rightarrow p_c^+$), but dimensional
analysis alone cannot determine the critical exponent $\nu$. The RG,
however, easily produces an asymptotic estimate, $\nu^{(n)}$, for a
given choice of $n = b^d$.  

Substituting an approximation of the correlation-length singularity,
\begin{equation}
\xi^{(n)}(p) \sim (p - p_c^{(n)})^{-\nu^{(n)}}, \label{eq:xipower}
\end{equation}
into Eq.~(\ref{eq:xisim}) with $m=n$ yields 
\begin{equation}
R_n(p) - p_c^{(n)} \sim n^{1/d\nu^{(n)}} (p - p_c^{(n)}),
\end{equation}
from which we identify, $n^{1/d\nu^{(n)}} =
R_n^\prime(p_c^{(n)})$. The resulting RG estimate of the critical
exponent is
\begin{equation}
\nu^{(n)} = \frac{1}{\log_b R_n^\prime(p_c^{(n)})}.  \label{eq:nu}
\end{equation}
(Note that, since $R_n(p)$ is generally a polynomial, it is analytic
at $p_c^{(n)}$, which implies that the correlation length diverges
with the same exponent for $p \rightarrow p_c^+$ and $p \rightarrow
p_c^-$.)  Such RG estimates of critical exponents are believed to
converge to their exact values in the large-cell limit, provided that
Eq.~(\ref{eq:Rlimit}) holds, consistent with our analysis below and
extensive numerical evidence in the literature.

\subsection{ The Renormalization Group }

Algebraically, the recursion in Eq.~(\ref{eq:pRG}) represents a group
of operators acting on the occupation probability (more precisely, a
semigroup without an inverse.) There are at least two equivalent ways
to construct this ``renormalization group''. First, by using the same
number of cells at each level of recursion, the renormalized
occupation probabilities, $p_{n^j} = \mathcal{R}^{(n)}_j p$, for the
sequence of system sizes, $N = n, n^2, n^3, n^4, \ldots$, can be
generated by iterates of the function, $R_n(p)$, acting on the site
occupation probability, $p \in [0,1]$. In this case, the group operators,
\begin{equation}
\mathcal{R}^{(n)}_j = R_n^{\circ j} = R_n \circ R_n \circ \ldots \circ
R_n \label{eq:ngroup}
\end{equation}
($j$ compositions of $R_n(p)$) are mappings from $[0,1]$ to $[0,1]$
which depend on the cell size, $n$.

Alternatively, one could view $R_n(p)$ as the initial condition for a
group of operators acting on the space of functions, $\{ f: [0,1]
\mapsto [0,1] \}$, by setting the number of cells to
equal the number of sites at each iteration.  For example, the
renormalized probabilities, $p_{n^{2^{j}}} = \mathcal{R}_j R_n(p)$,
for the sequence of system sizes, $N = n, n^2, n^4, n^8, \ldots$, can
be generated by group operators defined as follows,
\begin{eqnarray}
\mathcal{R}_{j+1} &=& \mathcal{R}_j
\mathcal{R}_1 \nonumber \\
\mathcal{R}_1 f &=& f\circ f .  \label{eq:group}
\end{eqnarray}
Although the initial condition depends on the function $R_n(p)$, in
this case the RG is always the same. As such, broad sets of initial
conditions can yield the same limiting behavior, and in this way the
RG explains, at least qualitatively, the universality of critical
exponents.

\subsection{ Cell Occupation by Spanning }
\label{sec:span}

These abstract concepts would not be terribly important were it not
for the fact that they can be put to practical use by clever
definitions of the cell occupation probability.  In the spirit of this
commemorative issue, we focus on the seminal contributions of
Reynolds, Stanley, and
Klein~\cite{reynolds77,reynolds78,reynolds80}. Before their work, a
few simple, small-cell decimation procedures were
proposed~\cite{young75,kirkpatrick77,yuge78a,yuge78b}, but these
authors first defined a cell to be ``occupied'' it is {\it spanned} by
a cluster from one side to another in at least one direction.  This
choice captures the essence of percolation as a {\it phase transition
in connectivity}. Moreover, it should always produce convergent RG
approximations because Eq.~(\ref{eq:Rlimit}) holds rigorously for any
spanning rule that, as $n \rightarrow \infty$, requires a cluster to
extend across an infinite distance~\cite{hughes}.

For a given choice of spanning rule, the spanning probability,
$R_n(p)$, can be evaluated exactly for small cells or approximately by
Monte-Carlo simulation for large
cells~\cite{reynolds77,reynolds78,reynolds80}.  For example, for a
square site or bond lattice in two dimensions, the function $R_n(p)$
could be the probability that a cluster connects opposite boundaries
of a square cell in one direction, both directions, or either
direction. Once $R_n(p)$ is obtained, estimates of $p_c$ and $\nu$ are
given by the expressions above.  The fractal dimension $D_f$ of
incipient infinite clusters can also be estimated by generalizing the
RG to include ``ghost bonds" between non-nearest neighbor sites,
although below we derive a different estimate which does not require
any ghost bonds.

For illustration purposes, throughout this paper we will consider the
following, very simple example.  For site percolation on the square
lattice, a reasonable RG for a small cell of only $n= 2\times 2$ sites
is based on the probability of spanning in one
direction~\cite{reynolds78,reynolds80},
\begin{equation}
R_4(p) = p^4 + 4 p^3 q + 2 p^2 q^2,   \label{eq:R4}
\end{equation}
(where $q=1-p$) which is the sum of probabilities to have spanning
clusters of size four, three, or two, respectively. Solving the
quartic equation, $R_4(p)=p$, yields (curiously) the golden mean as an
estimate of the critical point,
\begin{equation}
p_c^{(4)} = \frac{\sqrt{5} - 1}{2} \approx 0.618034,
\end{equation}
which is fairly close to the best available numerical
%
%
value~\cite{ziff00}, $p_c = 0.59274621(13)$. The critical exponent
estimate from Eq.~(\ref{eq:nu}),
\begin{equation}
\nu^{(4)} = \frac{1}{\log_2 \left[ 4 p_c^{(4)} ( 1 -
(p_c^{(4)})^2)\right]} \approx 1.63529,
\end{equation}
is not as close to the presumably exact value, $\nu=4/3$, although
larger cells slowly improve this estimate~\cite{reynolds80}. Of
course, more sophisticated and accurate RG schemes exist for various
percolation problems, but this simple example illustrates the power of
RG methods to approximate nontrivial quantities, such as the critical
exponents and $p_c$ itself, with remarkable ease. Using the same
example, we will produce RG estimates below for many other quantities,
both at and away from the critical point, with little additional
effort.

In the limit of large cells, $n \rightarrow\infty$, numerical evidence
suggests that RG estimates converge to their exact values (whenever
known), and the accuracy of small-cell approximations is often quite
remarkable~\cite{reynolds77,reynolds80,hovi96,hovi97,ziff02}. In spite
of these successes, however, the validity of the large-cell RG based
on spanning has recently been debated in the context of critical
spanning
probabilities~\cite{langlands92,grassberger92,ziff92,langlands94,stauffer94,grop94,hu94,hu95,hovi94,hovi96,lorenz98,ziff99,ziff02}.
Strictly speaking, Equation~(\ref{eq:pRG}) seems to predict that the
critical spanning probability in the infinite-system limit is
precisely equal to $p_c$.  On the other hand, simulations on the
square site lattice support another theoretical prediction (from
conformal field theory~\cite{cardy92}) that the critical spanning
probability in two dimensions is exactly $1/2$, which is somewhat less
than $p_c$.

There is no contradiction, however, since the RG only predicts that
the spanning probability at the $n$-cell estimate $p_c^{(n)}$ is equal
to $p_c^{(n)}$, which need not hold in the limit of infinite
cells~\cite{hovi96}. In fact, when the effect of the finite cell size
is properly considered as $p_c^{(n)} \rightarrow p_c$, the RG actually
predicts the universal spanning probability, including its subtle
dependence on the system
shape~\cite{langlands92,cardy92,hu95,lorenz98,ziff99} and
boundary conditions~\cite{grop94,hu94,ziff99} with good 
precision~\cite{hovi96,ziff02}.  In summary, the general validity of the RG
based on spanning is now fairly well established, and thus we use it as
a starting point for our theory below.

\subsection{ Self-Similarity Away From the Critical Point }
\label{sec:FSS}

The RG is often invoked to motivate (but not to calculate) the ``data
collapse" of any quantity, $Q(L,p)$ depending on system size $L$ and
concentration $p$ onto finite-size scaling functions of the form,
\begin{equation}
Q(L,p) \sim L^{D_f} \Phi\left( \frac{ L }{ \xi(p) } \right),  \label{eq:FSS}
\end{equation}
where the exponent, $D_f$, (which is the fractal dimension, in the
case of spanning clusters) and the dimensionless scaling function,
$\Phi(x)$, do not depend on $L$ or $p$. (Here and below, we always
choose units to set $a=1$.) As explained above, this result is simply
a consequence of dimensional analysis.

There is no existing theory to predict finite-size scaling
functions. Although the fractal dimension can be calculated from the
RG, the scaling function is usually obtained by the {\it ad hoc}
fitting of numerical
data~\cite{bunde,stauffer,landau-binder,bazant00}.  This approach,
however, is quite tedious and fails to take advantage of the full
power of the RG.

Although RG methods calculations have mostly been applied to the
critical phase, where $\xi \gg L$ (or $\xi = \infty$), the general
arguments given above only require that the correlation length be
large, $\xi \gg 1$, but not necessarily larger than the linear system
size $L$. Therefore, in principle {\it the RG can not only motivate,
but also predict, finite-size scaling functions,} as we demonstrate by
a suitable generalization below.

For RG approximations using cells of size $n = b^d$, the scaling law,
Eq.~(\ref{eq:FSS}), is replaced by
\begin{equation}
Q(L,p) \sim L^{D_f^{(n)}} \Phi^{(n)}\left( \frac{L}{\xi^{(n)}(p)}\right)
\end{equation}
where $\xi^{(n)}(p)$, $\Phi^{(n)}(x)$, $D_f^{(n)}$ are estimates based
on the $n$-cell level of coarse graining. The scaling function,
$\Phi^{(n)}(x)$, can be determined at any value of its argument,
$L/\xi^{(n)}(p)$, by an appropriate renormalization of $p$ to maintain
this ratio in the limit $N = L^d = n^j \rightarrow \infty$, as
explained below. Since $b/\xi(p) \rightarrow 0$ in this limit, the
general arguments above imply convergence, $\lim_{n\rightarrow\infty}
\Phi^{(n)}(x) = \Phi(x)$.

If the system size is large, $L \gg 1$, and one is not interested in
accurately describing crossover (which also requires being close to
the critical point, $\xi \gg 1$), then the RG can actually produce
accurate approximations far from the critical point, even when the
correlation length is small, $\xi(p) \approx 1$.  As emphasized in
Ref.~\cite{bazant00}, if a system is partitioned into cells with $b
\gg \xi$, then each cell makes independent contributions to any
quantity of interest for the whole system, which are easily combined
using standard techniques from probability theory.  The possibility of
accurate approximations in this limit is a consequence of
self-similarity near the stable, {\it off-critical} fixed points of the
RG, where $\xi \ll b \ll L$.

\section{ Stochastic Renormalization Group for Spanning-Cluster Masses }
\label{sec:SRG}

\subsection{ The Mathematical Statement of Statistical Self-Similarity }

A key ingredient missing in classical RG methods is the {\it
randomness of the order parameter in finite systems}. We now
reformulate the RG in probabilistic terms for the case of percolation,
using a general formalism which could also be applied to other
critical phenomena. Let $X_N(p)$ be an order parameter conditional on
spanning, such as the total mass of spanning clusters or the minimum
chemical distance between opposite boundaries, for a large system of
size $N\gg 1$ at a given concentration $p$, sufficiently close to the
critical point that discrete lattice effects are small, $\xi(p) \gg
1$. Consider coarse-graining such a system of size $N$ into $n$ cells
of size $m$.

The powerful constraint of statistical self-similarity discussed above
implies that, upon a suitable renormalization of the occupation
probability, the random number of occupied cells, $X_n(p_m)$, has the
same distribution as in a system of size $n$ with the original value
of $p$.  Unlike the sites or bonds in the original system, however,
the coarse-grained cells do not have equal masses. Instead, each
occupied cell contributes a random mass, $X_m(p)$, sampled
independently from the appropriate distribution for a smaller system
of size $m$.

This motivates a recursive construction we refer to as the
``Stochastic Renormalization Group'' (SRG): The total spanning-cluster
mass is given by a {\it recursive random sum of random variables},
\begin{equation}
X_{mn}(p) = \sum_{i = 1}^{ X_n(p_m) } X_m^{(i)}(p). \label{eq:XRG}
\end{equation}
where $X_m^{(i)}$ denotes the $i$th independent sample from the mass
distribution for each cell.  Random cluster masses, $X_n(p)$,
$X_{n^2}(p)$, $X_{n^3}(p)$, $\ldots$, at the $n$-cell level of
approximation can be generated recursively using Eq.~(\ref{eq:XRG}),
which is fully determined by the cell probability distribution,
\begin{equation}
f_n(p,x) = \mbox{Prob}(X_n(p) = x), \label{eq:fdef}
\end{equation}
and the cell occupation probability, $R_n(p)$, from
Eq.~(\ref{eq:pRG}).  These simple quantities can be either calculated
exactly for small cells or approximated numerically for large
cells. The SRG then extrapolates the distribution to larger finite
systems in an arbitrary stage of crossover between phases.

\begin{figure}
\begin{center}
\mbox{ \psfig{file=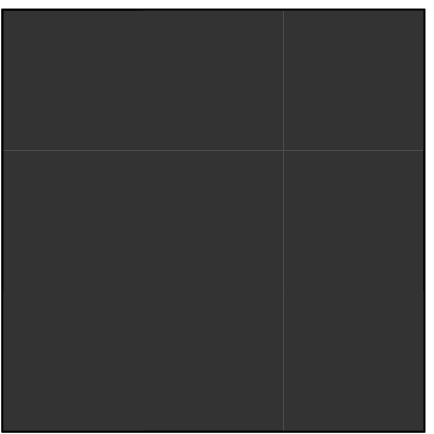,width=1.7in} \ \nolinebreak
\psfig{file=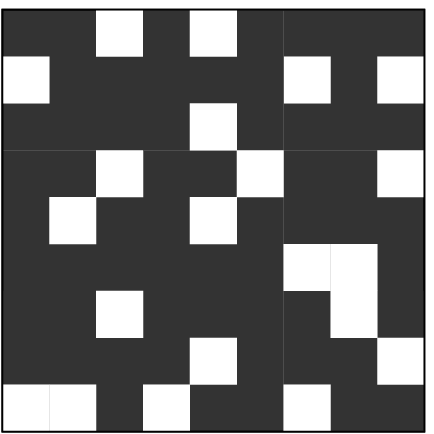,width=1.7in} \ \nolinebreak
\psfig{file=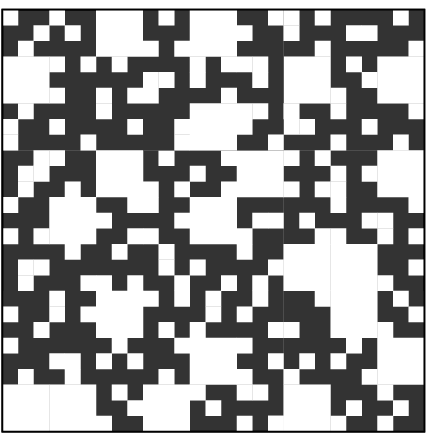,width=1.7in} }\\ \ \\ \mbox{
\psfig{file=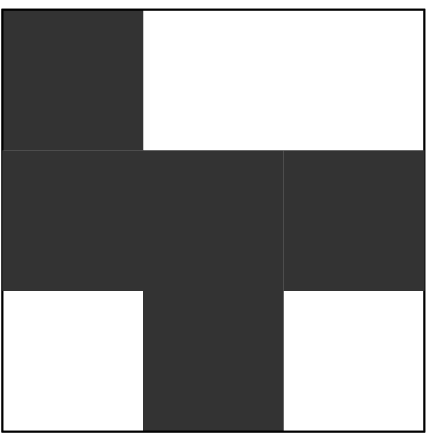,width=1.7in} \ \nolinebreak
\psfig{file=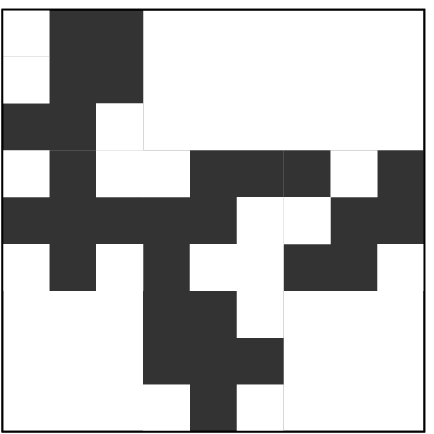,width=1.7in} \ \nolinebreak
\psfig{file=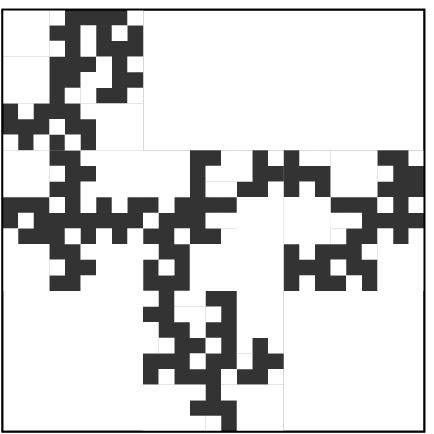,width=1.7in} }\\ \ \\ \mbox{
\psfig{file=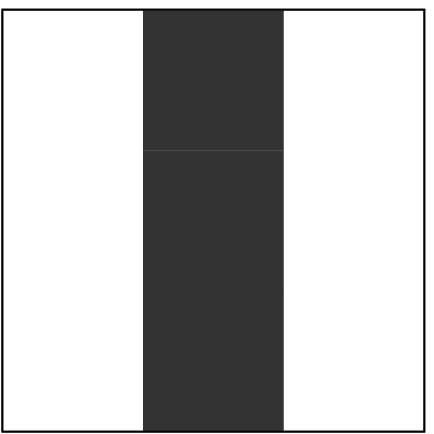,width=1.7in} \ \nolinebreak
\psfig{file=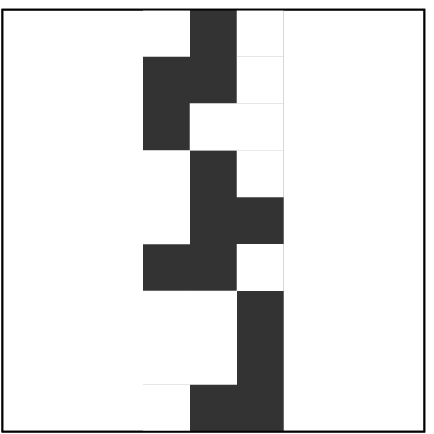,width=1.7in} \ \nolinebreak
\psfig{file=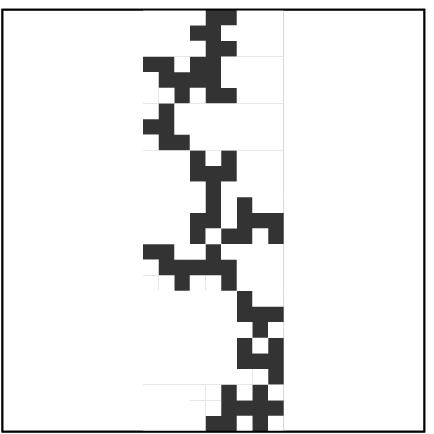,width=1.7in} }
\begin{minipage}[h]{5.5in}
\caption{ \label{fig:spanning_clusters} Three random realizations (in
each row from left to right) of a nine-cell SRG branching process for
the total cluster mass conditional on spanning in the vertical
direction. Top row: the supercritical phase, showing the crossover to
critical behavior at scales smaller than the correlation
length. Middle row: the critical phase, showing complete statistical
self-similarity.  Bottom row: the subcritical phase, also showing the
crossover to critical behavior. (Note that in these examples the
fractal dimension has been reduced well below its actual value for $2d$
percolation, $D_f = 91/48$, to illustrate the crossover regime with
greater clarity.)}
\end{minipage}
\end{center}
\end{figure}

The SRG approximates the statistics of cluster masses through random
geometrical constructions, as illustrated in
Figure~\ref{fig:spanning_clusters}.  The coarse-graining procedure
described above involves discarding contributions from unoccupied
cells, and the figure only shows the sites and cells which survive the
entire process (from right to left).  Equivalently, clusters are
created by recursively refining occupied cells (from left to
right). In each row, the picture on the right shows a random cluster of
sites, and the pictures to the left show the hierarchical structure of
connectivity among the sites at different scales of
coarse-graining. At scales smaller than the correlation length, these
artificially generated clusters are random fractals characteristic of
the critical phase, while at larger scales in normal phases they are
objects of integer dimension, as explained below.  Although the
connectivity of these artificial clusters is not required explicitly
(as shown in the figure), their mass distributions must nevertheless
approach those of the real spanning clusters in each phase in the
large-cell limit, purely as a consequence of statistical
self-similarity. The crossover between phases can also be described if
the correlation length is much larger than lattice spacing.

\subsection{ Analogy with Population Growth }

In mathematical terms, Equation~(\ref{eq:XRG}) describes a stochastic
{\it branching process}~\cite{feller,harris,athreya}, which is notably
non-stationary since the generating distribution, $f_n(p_m,x)$, is
different at every stage of recursion.  Since the nineteenth century,
stationary branching processes have been analyzed as models of
population growth, e.g. in the contexts of human family
trees~\cite{watson1874}, animal populations~\cite{fisher30} and
nuclear chain reactions~\cite{schroedinger}. It does not appear,
however, that any mathematical studies have addressed the particular
type of non-stationarity in Eq.~(\ref{eq:XRG}), which has multiple
fixed points corresponding to different thermodynamic phases.

The analogy with population growth provides a delightfully simple way
to understand the SRG: {\it The hierarchical structure of power-law
correlations is analogous to the genealogical structure of
populations}. This connection is easily understood from the graphical
constructions of Fig.~\ref{fig:spanning_clusters}. Each cluster of
sites is a graphical representation of a ``family tree''. The
``ancestors'' of a site are the coarse-grained cells in which it is
contained.  From the pictures, it is clear that any random changes to
the occupied cells (``ancestors'') at a given length scale (or
``generation'') greatly affect the number of occupied cells and sites
(``descendants'') at all smaller length scales.  Therefore, even
though all of the random variables in the branching process are
independent, long-range spatial correlations, analogous to ``family
ties'', are induced on the final ``population'' of sites.

Although the analogy with population growth is conceptually very
useful, the SRG branching process involves several new features not
typically studied in the context of populations, and thus it requires
new mathematical results. For example, the large-cell limit (of
infinitely many ``children'' per individual) is of primary interest
for the SRG. More importantly, the structure of spatial correlations
would corresponds to a very unusual ``family tree'' where randomness
in the number of children only appears gradually after a certain
generation. In the SRG, this occurs when the cell size is comparable
to the correlation length (the middle column in
Fig.~\ref{fig:spanning_clusters}). Below this scale, spatial
correlations induced on the sites decay as a power law, reflecting the
absence of a characteristic length, while above this scale
correlations decay exponentially because the number of ``children'' is
no longer random. This gradual variation in the probability distribution
for ``births'' in the branching process causes a phase transition as
the linear system size passes through the correlation length, as
explained below.


\subsection{ Recursions for Distributions and Generating Functions }

Since all the random variables in the SRG branching process are
independent, Equation~(\ref{eq:XRG}) takes the form~\cite{feller},
\begin{equation}
f_{mn}(p,x) = \sum_{j=x_{n}^-}^{x_{n}^+} f_n(p_m,j) f_m^{\ast
j}(p,x), \label{eq:fRG}
\end{equation}
in terms of probability distribution functions, where $f^{\ast j} = f
\ast f \ast \ldots \ast f$ ($j$ times) denotes the $j$th
auto-convolution, defined by $f \ast g(x) = \sum_y f(x-y)g(y)$.  In
general, the bounds of summation, $x_n^-$ and $x_n^+$, are set by the
distribution function, $f_n(p,x)$, which also initializes the
recursion in Eq.~(\ref{eq:fRG}). In the case of percolation quantities
subject to a spanning condition, the minimum number of occupied cells
is of the order of the linear cell size, $x_n^- = O(b)$,
e.g. $x_n^- = b = n^{1/d}$ for hypercubic cells, and the maximum
number occupied cells is always equal to the total number of cells,
$x_n^+ = n$.

For both analytical and numerical calculations with the SRG, it is
convenient to work with various generating functions~\cite{feller},
which are essentially discrete Fourier or Laplace transforms. For
example, Equation~(\ref{eq:fRG}) takes the simple form,
\begin{equation}
\bar{f}_{mn}(p,z) = \bar{f}_n(p_m,\bar{f}_m(p,z)),
\end{equation}
in terms of probability generating functions,
\begin{equation}
\bar{f}(z) = \sum_{x=0}^\infty f(x) z^x.
\end{equation}
Other useful forms of the SRG,
\begin{eqnarray}
\tilde{f}_{mn}(s) &=& \tilde{f}_m(p_m,\log \tilde{f}_m(p,s)) \\
\hat{f}_{mn}(p,s) &=& \hat{f}_n(p_m,\hat{f}_m(p,s)), \label{eq:fhatRG}
\end{eqnarray}
are expressed in terms of moment generating functions,
\begin{equation}
\tilde{f}(s) = \bar{f}(e^s),
\end{equation}
or cumulant generating functions, 
\begin{equation}
\hat{f}(s) = \log \tilde{f}(s),
\end{equation}
respectively.  These transformations reduce the complicated sum of
convolutions in Eq.~(\ref{eq:fRG}) to a simple composition of
generating functions.  In physics parlance, this amounts to changing
the representation of the SRG from real space to various definitions
of reciprocal (or momentum) space.

\subsection{ The Stochastic Renormalization Group }

Let us consider the algebraic structure of the transformations
involved in the branching process, Eq.~(\ref{eq:XRG}).  There are
various ways to iterate the recursions above, as we now illustrate
using the cumulant-generating-function representation,
Eq.~(\ref{eq:fhatRG}). For example, as in Eq.~(\ref{eq:ngroup}) the
same number of cells $n$ can be used at each level of recursion, which
yields approximations,
\begin{equation}
\hat{f}_{n^j}(p) = \hat{f}_n(p_{n^{j-1}}) \circ \hat{f}_n(p_{n^{j-2}}) \circ
\ldots \circ \hat{f}_n(p_n) \circ \hat{f}_n(p),
\end{equation}
 for the sequence of system sizes, $N = n, n^2, n^3, n^4, \ldots$.
Alternatively, as in Eq.~(\ref{eq:group}) the number of cells can be
set equal to the number of sites at each level of recursion, which
yields the (identical) approximations
\begin{eqnarray*}
\hat{f}_{n^2}(p) &=& \hat{f}_n(p_n) \circ \hat{f}_n(p) \\
\hat{f}_{n^4}(p) &=& \hat{f}_{n^2}(p_{n^2}) \circ \hat{f}_{n^2}(p)  \\
  &=& \hat{f}_n(p_{n^3}) \circ \hat{f}_n(p_{n^2}) \circ \hat{f}_n(p_n)
\circ \hat{f}_n(p) \\
& \vdots &  \\
\hat{f}_{n^{2^{j+1}}}(p) &=& \hat{f}_{n^{2^j}}(p_{n^{2^j}})
\circ \hat{f}_{n^{2^j}}(p) \\
&=& \hat{f}_n(p_{n^{2^{j+1}}-1}) \circ \hat{f}_n(p_{n^{2^{j+1}}-2}) \circ
\ldots \circ \hat{f}_n(p_n)
\circ \hat{f}_n(p) 
\end{eqnarray*}
for the sequence of system sizes, $N = n, n^2, n^4, n^8, \ldots$.  At
a fixed point, the same set of RG operators $\{
\mathcal{R}_j \}$ as in Eq.~(\ref{eq:group}) now acts on initial
conditions, $\hat{f}_n(p,s)$, in the space of functions, $\{
f:[0,1]\times \mathcal{C} \mapsto \mathcal{C} \}$.

If $p$ is not a fixed point, then the functional transformations are
somewhat more complicated, but they still have the structure of an
Abelian semigroup (hence the name ``Stochastic Renormalization
Group'') in terms of a set of operators, $\{ \mathcal{R}^{SRG}_j \}$,
defined by
\begin{eqnarray}
\mathcal{R}_{j+1} &=& \mathcal{R}^{SRG}_1
\mathcal{R}^{SRG}_j, \nonumber \\
\mathcal{R}^{SRG}_1 \cdot \{ R, f(p) \} &=& \{ R\circ R, f(R(p))
\circ f(p) \} .
\end{eqnarray}
These functional operators act on initial conditions, $\{ R_n(p),
\hat{f}_{n}(p,s) \}$, in the function space, $\{ R:[0,1]\mapsto[0,1]\}
\times \{ \hat{f}:[0,1]\times \mathcal{C} \mapsto \mathcal{C} \}$,
\begin{equation}
\{  R_{n^{2^j}}(p), \hat{f}_{n^{2^j}}(p,s) \} = \mathcal{R}^{SRG}_j \{
R_n(p), \hat{f}_{n}(p,s) \},
\end{equation}
where $R_n(p)$ is the spanning probability and $\hat{f}_{n}(p,s)$ is
the cumulant generating function for the spanning cluster mass in a
cell of size $n$.

The fact that the SRG can be expressed as a {\it universal} group
acting on a space of initial conditions suggests that the crossover of
distributions between different phases shares the same universality
properties as the fixed points. Therefore, the shapes of scaling
functions and the limiting distributions in the critical and crossover
regimes should depend on embedding dimension, but not on the
microscopic details of the lattice (although we do not claim to
rigorously prove this universality here, even within the SRG
formalism).  
Below, we will argue that boundary conditions also affect
the universal limiting distributions and thus also the scaling
functions for crossover.

\section{ A Theory of Crossover Phenomena }
\label{sec:cross}

\subsection{ Two Types of Fluctuations }

It is straightforward to derive recursions for the moments,
\begin{equation}
\mu_{n,k}(p) = \langle X_n(p)^k \rangle =
\left(z\frac{\partial}{\partial z}\right)^k\bar{f}_n(p,1) = \left(
\frac{\partial}{\partial s}\right)^k \tilde{f}_n(p,0),
\end{equation}
or the cumulants,
\begin{equation}
c_{n,k}(p) =  \left( \frac{\partial}{\partial s}\right)^k \hat{f}_n(p,0),
\label{eq:cumdef}
\end{equation}
by differentiating the various generating-function recursions
above. For example, the mean, $\mu_n(p) = \mu_{n,1}(p) = c_{n,1}(p)$,
satisfies the recursion,
\begin{equation}
\mu_{mn}(p) = \mu_n(p_m) \mu_m(p),
\label{eq:muRG}
\end{equation}
which simply confirms our intuition that the mean total mass is
equal to the mean number of occupied cells times the mean mass per cell.
The variance, $\sigma_n(p)^2 = c_{n,2}(p)$, satisfies another recursion,
\begin{equation}
\sigma_{mn}(p)^2 = \mu_n(p_m) \sigma_m(p)^2 + \sigma_n(p_m)^2 \mu_m(p)^2,
\label{eq:sigRG}
\end{equation}
which is coupled to that of the mean.

The recursion for the variance, Eq.~(\ref{eq:sigRG}), has an important
physical interpretation.  The first term on the right side is the
variance within each cell times the mean number of occupied
cells. This represents the well-known additivity of variance for sums
of a fixed number of independent random variables.  The second term is
the {\it additional variance due to fluctuations in the number of
occupied cells}. As explained above, this kind of fluctuation induces
hierarchical (power-law) correlations between the microscopic site
variables. The gradual disappearance of this second source of variance
at scales larger than the correlation length is precisely what
determines the crossover away the critical phase.

The higher cumulants are important indications of this crossover since
they measure departures from a Gaussian distribution, so we give a few
more recursions implied by the SRG branching process. The skewness,
$c_{N,3}(p)$, satisfies,
\begin{equation}
c_{mn,3}(p) = \mu_n(p_m) c_{m,3}(p) + 3 \sigma_n(p_m)^2 \mu_m(p)
\sigma_m(p)^2 + c_{n,3}(p_m) \mu_m(p)^3 ,
\end{equation}
and the kurtosis, $c_{N,4}(p)$, satisfies,
\begin{eqnarray}
c_{mn,4}(p) &=& \mu_n(p_m) c_{m,4}(p) + \sigma_n(p_m)^2 \left( 4 \mu_m(p)
c_{m,3}(p)  +  3 \sigma_m(p)^4 \right) \nonumber \\
& &  + 3 c_{n,3}(p_m) \mu_m(p)^2
\sigma_m(p)^2  +  c_{n,4}(p_m) \mu_m(p)^4 .  \label{eq:c4RG}
\end{eqnarray}
On the right side of each of these recursions, the first term
represents the additivity of the cumulants of the mass within each
cell over the expected number of occupied cells, as in sums of
independent variables. The remaining terms represent additional
contributions due to fluctuations in the number of occupied cells. As
we elucidate below, {\it the phase transition occurs as a result of
the competition between these two kinds of fluctuations, representing
short-range and long-range correlations, respectively}.

\subsection{ Crossover of Probability Distributions }

As discussed in section~\ref{sec:RG}, large systems near the critical
point which have the same ratio, $L/\xi(p)$, are in the same stage of
crossover, i.e. they are statistically identical. This situation can
be conveniently arranged within the formalism of the SRG, where the
analogous ratio is $L/\xi^{(n)}(p)$: {\it Different systems are in the
same stage of crossover if they have the same cell occupation
probability on the largest cells}, i.e. at the highest level of
coarse-graining.  In particular, a sequence of systems of increasing
size can be frozen in the same state of crossover by adjusting $p$ at
each stage of recursion to set the occupation probability on the
largest cells to a given constant, $0 < \eta < 1$, which acts as the
finite-size scaling variable. 

For example, if the cells have reached size $m = n^j$ for a system of
size $N=n^{j+1}$, then we adjust $p$ to the following value
\begin{equation}
p = R_n^{-j}(\eta) =
R_n^{-1}\circ R_N^{-1}\circ \ldots R_N^{-1}(\eta)
\end{equation}
(the inverse of $R_n(p)$ applied $j$ times).  This choice ensures that
the correlation length for the system is $\xi^{(n)}(\eta)$ in
units of the largest cell length.  In other words, the finite-size
scaling variable is held constant,
\begin{equation}
\frac{\xi^{(n)}(p)}{L} = \frac{\xi^{(n)}(\eta) n^{j/d}}{L} =
\frac{\xi^{(n)}(\eta)}{b},  \label{eq:eta}
\end{equation}
where $n = b^d$ and $N = L^d$. 

The crossover from subcritical to critical to supercritical is
controlled by varying $\eta$ from 0 to $p_c^{(n)}$ to 1, respectively.
For a given value of $\eta$, if we define,
\begin{equation}
\hat{h}_{n^{j+1}}(\eta,s) =
\hat{f}_{n^j}(R_n^{-j}(\eta),s),
\end{equation}
then the SRG 
takes the form,
\begin{eqnarray}
\hat{h}_n(\eta) &=& \hat{f}_n(\eta) \nonumber \\ \hat{h}_{n^2}(\eta)
&=& \hat{f}_n(\eta) \circ \hat{f}_n(R_n^{-1}(\eta)) \nonumber \\
\hat{h}_{n^3}(\eta) &=& \hat{f}_n(\eta) \circ
\hat{f}_n(R_n^{-1}(\eta)) \circ \hat{f}_n(R_n^{-2}(\eta)) \nonumber \\
&\vdots & \nonumber \\ \hat{h}_{n^{j+1}}(\eta) &=& \hat{f}_n(\eta)
\circ \hat{f}_n(R_n^{-1}(\eta)) \circ \ldots \circ
\hat{f}_n(R_n^{-j}(\eta))  \label{eq:hhat}
\end{eqnarray}
in the  cumulant-generating-function representation.  By construction,
the inverse  iteration of the occupation probability  converges to the
critical fixed point,
\begin{equation}
\lim_{j\rightarrow\infty}
R_n^{-j}(\eta) = p_c^{(n)}
\end{equation}
for $0 < \eta < 1$, so the iteration for $\hat{h}_N(\eta,s)$
approaches a stationary functional iteration, which has a well defined
limit (after an appropriate rescaling, discussed below).  If we
next take the large-cell limit, $n\rightarrow\infty$, then
convergence is attained because the first term in the resulting
infinite product tends to the true distribution inside a large cell of
size $n \rightarrow \infty$, while each of the remaining terms tends
toward a constant, $\hat{h}_n(p_c^{(n)},s)$, because (again, by
construction) $\lim_{n\rightarrow\infty} R_n^{-j}(\eta) =
p_c^{(n)}$. Therefore, in the large-cell limit
{\it the SRG produces asymptotic approximations of scaling functions
and probability distributions in the crossover regime}.

\subsection{ Finite-Size Scaling Functions }

An analytical formula for the scaling function of the $k$th cumulant,
\begin{equation}
c_{L^d,k}(p) \sim (L^{D_f^{(n)}})^k 
\Phi^{(n)}_k\left(\frac{L}{\xi^{(n)}(p)}\right), \label{eq:Phik}
\end{equation}
can be derived from the $n$-cell SRG. The trick is to choose $\eta$
according to Eq.~(\ref{eq:eta}) and thus cast the scaling function in
the equivalent form,
\begin{equation}
\Psi^{(n)}_k(\eta) =
\Phi^{(n)}_k\left(\frac{b}{\xi^{(n)}(\eta)}\right).  \label{eq:Psidef}
\end{equation}
Although the subcritical ($p < p_c^{(n)}$) and supercritical ($p >
p_c^{(n)}$) scaling functions are different in terms of the variable
$x= L/\xi^{(n)}(p)$, there is a single, continuous scaling function in
terms of the variable $\eta\in [0,1]$.  From the auxiliary function,
$\Psi^{(n)}_k(\eta)$, the usual scaling function for the $k$th moment,
$\Phi^{(n)}_k(x)$, can be determined from Eq.~(\ref{eq:Psidef}), once
the appropriate $n$-cell correlation length, $\xi^{(n)}(p)$, is
measured (e.g. as in Fig.~\ref{fig:pn}).

The auxiliary scaling function, $\Psi^{(n)}_k(\eta)$, is easily
calculated (at least numerically) from Eq.~(\ref{eq:hhat}). For
example, consider the case of the mean, $k=1$. After $j$ iterations of
the SRG, an approximation of the scaling function,
$\Psi_1^{(n)}(\eta)$, is given by 
\begin{equation}
\frac{\mu_{n^{j+1}}(p)}{\mu_n(p_c^{(n)})^{j+1}}
= \frac{\mu_n(\eta)}{\mu_n(p_c^{(n)})} \cdot 
 \frac{\mu_n(R_n^{-1}(\eta))}{\mu_n(p_c^{(n)})} \cdot
\frac{\mu_n(R_n^{-2}(\eta))}{\mu_n(p_c^{(n)})}  
\cdot \ldots \cdot
 \frac{\mu_n(R_n^{-j}(\eta))}{\mu_n(p_c^{(n)})},  \label{eq:Psij}
\end{equation}
where the rescaling factors in the denominator are chosen so that
\begin{equation}
\Psi_{1}^{(n)}(p_c^{(n)}) = \Phi_1^{(n)}(0) = 1.
\end{equation}
This is also required for convergence as $j \rightarrow\infty$ (with
$\eta$ fixed) since $R_n^{-j}(\eta) \rightarrow p_c^{(n)}$.  In this
limit, we obtain a formula for the $n$-cell SRG approximation of the
infinite-system scaling function,
\begin{equation}
\Psi^{(n)}_1(\eta) = 
\prod_{j=0}^\infty \frac{\mu_n(R_n^{-j}(\eta))}{\mu_n(p_c^{(n)})},
\label{eq:Psiprod}
\end{equation}
It is straightforward to prove that this infinite product converges if
$\mu_n^\prime(p_c^{(n)}) > 0$ and $R_n^\prime(p_c^{(n)}) > 0$, as
required for the critical exponents to be finite and nonzero.  

By comparing Eq.~(\ref{eq:Phik}) and Eq.~(\ref{eq:Psij}), we have 
\begin{equation}
L^{D_f^{(n)}} = n^{D_f^{(n)}j/d} = \mu_n(p_c^{(n)})^j
\end{equation}
which yields a simple estimate of the fractal dimension,
\begin{equation}
D_f^{(n)} = d\; \log_n \mu_n(p_c^{(n)}). \label{eq:Df}
\end{equation}
It is apparent that this estimate of the fractal dimension converges
to the correct value in the large-cell limit, as long as $p_c^{(n)}
\rightarrow p_c$, because it is equivalent to the usual finite-system
estimate used in numerical simulations, $\mu_n(p_c) \sim n^{D_f/d}$.
Equation (\ref{eq:Df}) has also been derived by Hovi and Aharony using
similar methods, albeit valid only at the critical point~\cite{hovi97}
(see also below). Here, however, we also predict scaling functions and
probability distributions for the complete crossover between phases.

As $n \rightarrow \infty$, the first term in the infinite product of
Eq.~(\ref{eq:Psiprod}) converges to the true scaling function, and all
other terms converge to unity, since $\lim_{n\rightarrow\infty}
R_n^{-1}(\eta) = p_c^{(n)}$ for $0 < \eta < 1$ according to
Eq.~(\ref{eq:Rlimit}). This suggests that the SRG approximation of the
scaling function converges in the large-cell limit. A very similar
convergence proof is carried out in detail in Part II for an
infinite-product formula for the strength of the infinite cluster.

\subsection{ A Simple Example }

To illustrate the SRG approach, we compute estimates of finite-size
scaling functions for the total mass of clusters spanning a square
system in one direction, using the very simple four-cell scheme
introduced above in section~\ref{sec:span}. In this case, the SRG is
based on the conditional probability generating function,
\begin{equation}
\bar{f}_4(p,z) = \frac{ 2 p^2 q^2 z^2 + 4 p^3 q z^3 + p^4 z^4}{
R_4(p)},  \label{eq:f4}
\end{equation}
where $R_4(p)$ is defined in Eq.~(\ref{eq:R4}), which yields the
following polynomials for the moments,
\begin{equation}
\mu_{4,k}(p) = \frac{2^{k+1}  p^2 q^2 + 3^k 4 p^3 q z^3 + 4^k
p^4}{R_4(p)}. 
\label{eq:mu4} 
\end{equation}
From these expressions it is straightforward to compute (at least
numerically) the auxiliary scaling functions $\Psi^{(n)}_k(\eta)$, but
it would also be necessary to calculate the correlation length,
$\xi^{(n)}(p)$ to obtain the usual scaling functions, $\Phi_k^{(n)}(x)$.

\begin{figure}
\begin{center}
\mbox{ 
\psfig{file=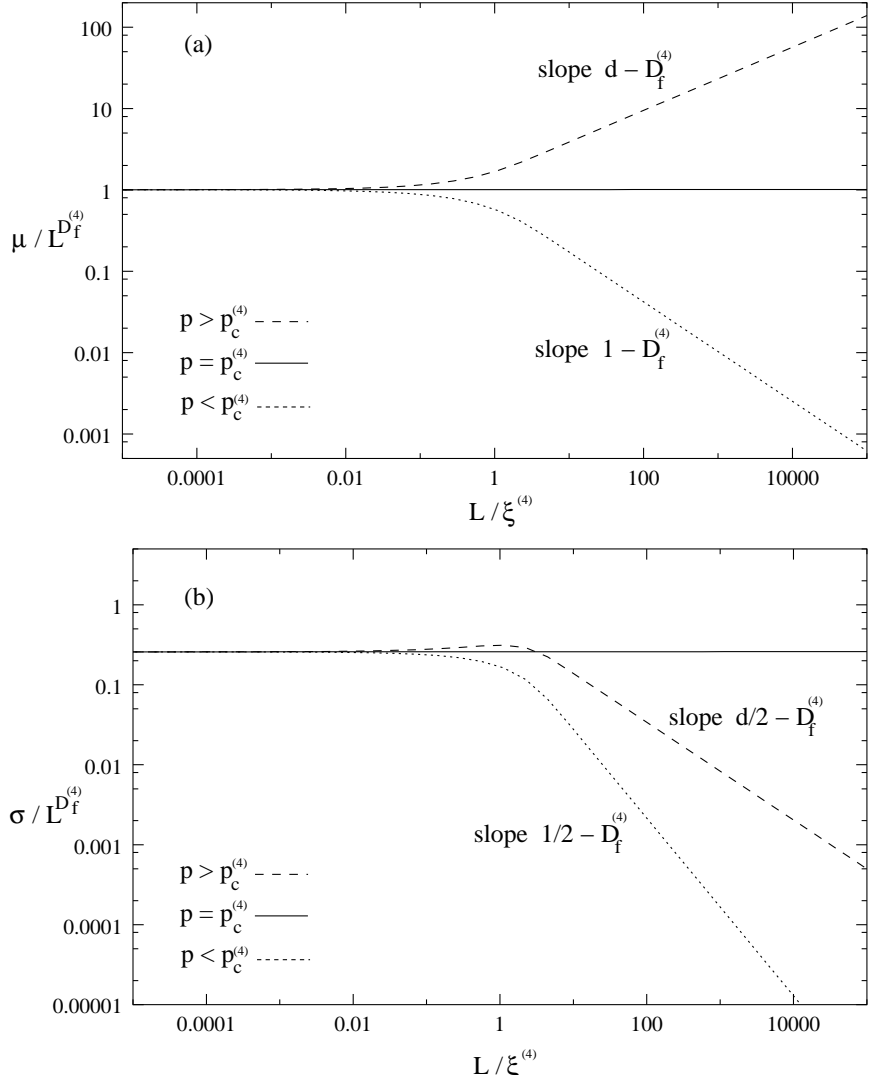,width=4.5in}
}
\begin{minipage}[h]{5.5in}
\caption{ \label{fig:crossover} Four-cell SRG estimates of
crossover functions for (a) the mean and (b) the standard deviation of
the total mass of clusters spanning a square system in one
direction. }
\end{minipage}
\end{center}
\end{figure}

In practice, it is easier to iterate the cumulant recursions
(numerically) starting from two particular values of $p$ very close to
$p_c^{(n)}$ (one above and one below), as in Fig.~\ref{fig:pn}, and
then rescale by appropriate values of $\xi^{(n)}(p)$ for the chosen
values of $p$. This method effectively samples the true continuous
scaling functions at discrete points $b^j/\xi^{(n)}(p)$, which are
very closely spaced (in log coordinates). For our four-cell example,
the results of this method are shown in Fig.~\ref{fig:crossover}.

The scaling functions in Fig.~\ref{fig:crossover} exhibit a smooth
crossover from one power law to another (straight lines on a log-log
plot) as the system size exceeds the correlation length, which is the
correct qualitative behavior.  The quantitative accuracy of these
particular approximations is not very good, however, because even the
fractal dimension (limiting slope) is poorly described. For this
example, Equation~(\ref{eq:Df}) predicts $D_f^{(n)} \approx 1.611$,
for the mass of spanning clusters, compared to the presumably exact
value, $D_f = 91/48 \approx 1.896$.  

As with the critical exponent,
$\nu$, four cells ($b=2$) is not enough to reasonably approximate the
fractal dimension, $D_f$, but these estimates improve very quickly
with increasing cell size.  For example, Equation~(\ref{eq:Df})
predicts fractal dimensions for various structural quantities on the
square bond lattice with less one percent error using remarkably small
cells ($b\leq 4$)~\cite{hovi97}.  Likewise, preliminary numerical
results with the SRG using larger cells on the square site lattice
also suggest that the approximation of scaling functions improves
quickly with cell size, concomitant with the improvement in describing
the fractal dimension. This ongoing work will be reported elsewhere,
but at least the analysis here predicts convergence in the large-cell
limit (provided that the scaling functions actually exist).

\section{ The Supercritical Fixed Point }
\label{sec:CLT}

\subsection{ Asymptotic Independence }

A detailed analysis of the fixed points of the SRG will be presented
in upcoming parts of this series, so here in the remaining sections we
simply give the basic ideas.  

Above the critical point of the RG, $p > p_c^{(n)}$, once the linear
cell size exceeds the correlation length, $m^{1/d} \gg \xi^{(n)}(p)$,
the renormalized cell occupation probability approaches the
supercritical fixed point, $p_m \rightarrow 1$.  In this limit, the
sums in Eq.~(\ref{eq:XRG}) no longer have a random number of terms
because $X_n(1)$ is equal to a constant, $x_n^+$, with probability
one, i.e. 
\begin{equation}
\bar{f}_n(1,z) = z^{x_n^+}, \ \ \ \tilde{f}_n(1,s) = e^{x_n^+ s},
\mbox{ \ and \ } \hat{f}_n(1,s) = x_n^+ s.  \label{eq:supgen}
\end{equation}
The suppression of randomness in the number of ``children'' of the SRG
branching process corresponds to the removal of hierarchical
correlations at scales larger than the correlation length. In the
infinite-system limit, independent contributions of mass from all
cells are simply accumulated at each level of recursion. In other
words, the SRG at the supercritical fixed point reduces to a {\it
non-random sum of independent random variables},
\begin{equation}
X_{mn}(p) \sim \sum_{i=1}^{x_n^+} X_m^{(i)}(p).  \label{eq:XRGsuper}
\end{equation}
For example, at scales larger than the correlation length in the
supercritical phase, all $x_n^+ = n = b^d$ cells make independent
contributions to the total mass of spanning clusters, while only a
straight line of $x_n^+ = b$ cells contributes to the minimum chemical
distance in a hypercubic system. (The former case is illustrated in
the top row of Figure~\ref{fig:spanning_clusters}.)

\begin{figure}
\begin{center}
\mbox{ 
\psfig{file=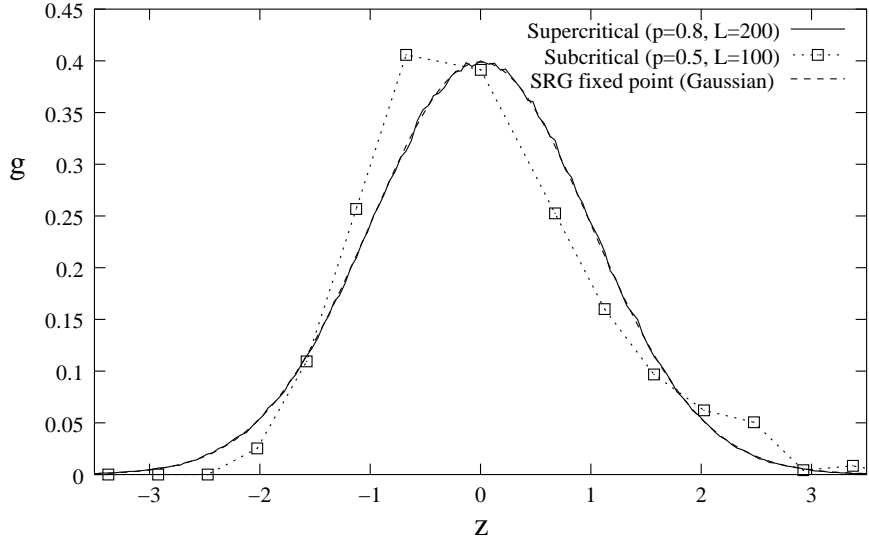,width=4.5in}
}
\begin{minipage}[h]{5.5in}
\caption{ \label{fig:spanxory_offcrit} Supercritical and subcritical
probability distributions, $g(z)$, for the total mass of clusters
spanning a square system in either direction (scaled to have zero mean
and unit variance). Numerical simulation data for square-lattice site
percolation is compared with the analytical prediction of the SRG at
the off-critical fixed points, i.e. the Gaussian,
$e^{-z^2/2}/\sqrt{2\pi}$. The numerical data reflects averaging over
10 million samples for a system of size $N = 200 \times 200$ for the
supercritical case and 20 million samples of size $N = 100 \times 100$
for the subcritical case.  }
\end{minipage}
\end{center}
\end{figure}

\subsection{ Central Limit Theorem }

It is well known that non-random sums of independent random variables
obey the Central Limit Theorem (CLT), which is characterized by a
Gaussian limiting distribution with linear scaling of the mean and
variance~\cite{feller}. In the case of the SRG, ``the CLT holds''
means that the distribution of the scaled random variable 
\begin{equation}
Z_N = \frac{X_N(p) -
\mu_N(p)}{\sigma_N(p)}
\end{equation} 
converges to a standard normal distribution (i.e. a Gaussian with zero
mean and unit variance). If equality holds in Eq.~(\ref{eq:XRGsuper})
and thus Eq.~(\ref{eq:supgen}) applies, then the CLT is
straightforward to prove from the cumulant generating function
representation,
\begin{equation}
\hat{f}_{mn}(s,p) = x_+ \hat{f}_m(s,p),
\end{equation}
which implies the additivity of cumulants (the crucial step in most
proofs of the CLT). This recursion is only asymptotically valid,
however, so it is not clear {\it a priori} that the convergence is
fast enough to ensure that the CLT holds.

Nevertheless, in Part II it is shown that the CLT does in fact hold
for the supercritical fixed point of the SRG. This result is
consistent with Newman's proofs of the CLT for the mass of finite
fragments of the infinite
cluster~\cite{newman80,newman81b,newman83}. The SRG prediction is also
in excellent agreement with our own new simulation results (using the
methods of Ref.~\cite{bazant00}) for the total mass of clusters
spanning a large square system in either direction, shown in
Fig. ~\ref{fig:spanxory_offcrit}.

\section{ The Subcritical Fixed Point }

For $p < p_c^{(n)}$, the occupation probability renormalizes to zero
because the spanning probability tends to zero (exponentially) once
the cell size exceeds the correlation length. In this limit, all cells
become ``unoccupied'', and hence spanning clusters are extremely
rare. When spanning clusters do occur, however, they tend to include
the minimum number of cells, $x_n^-$, in a straight line across the
system ($x_n^- = b$), as illustrated in
Fig.~\ref{fig:spanning_clusters}. Indeed, this intuition is supported
by the SRG for the {\it conditional} distribution of spanning cluster
masses: At the subcritical fixed point, the SRG again reduces to a
non-random sum of random variables,
\begin{equation}
X_{mn}(p) \sim \sum_{i=1}^{x_n^-} X_m^{(i)}(p),  \label{eq:XRGsub}
\end{equation}
since $X_n(0) = x_n^-$ with probability one. 

As in the supercritical case, this suggests that the CLT holds. The
simulation data in Fig.~\ref{fig:spanxory_offcrit} is also consistent
with this claim, although it is difficult to generate enough
subcritical spanning clusters to attain good convergence of the
distribution. (Much better convergence could presumably be achieved
using a ``go-with-the-winners'' algorithm~\cite{grassberger00}.)  To
the author's knowledge, the validity of the CLT for various quantities
conditional on spanning in subcritical percolation has not previously
been reported.

Since spanning clusters are very rare in the subcritical regime, a
much better order parameter for the entire phase transition in finite
systems is the mass of the largest
cluster~\cite{stauffer80,family80,margolina84,duxbury87,bazant00}. At
the subcritical fixed point, the mass distribution of the largest
cluster does not obey the CLT, and instead the Fisher-Tippett limit
theorem of extreme statistics holds~\cite{duxbury87,bazant00}. On the
other hand, the CLT does hold for the largest-cluster mass at the
supercritical fixed point because the largest cluster and various
spanning clusters are all the same in the supercritical limit. To
unify the present theory with these intriguing results revealing
subcritical/supercritical asymmetry, in Part IV the SRG is extended to
the more subtle case of the mass of the largest cluster
(which may not span).

\section{ The Critical Fixed Point }
\label{sec:FCLT}

\subsection{ Complete Statistical Self-Similarity }

At the critical fixed point, $p = p_c^{(n)}$, the SRG reduces
exactly to 
\begin{equation}
X_{mn} = \sum_{i=1}^{X_n} X_m^{(i)}, \label{eq:XRGc}
\end{equation}
which asserts that, at every scale, the random cluster mass $X_{mn}$
in a system of $N=mn$ sites has the same statistics as the sum of
independent mass contributions $X_m^{(i)}$ from a random number $X_n$
out of $n$ cells of size $m$.  (In this section, we will drop the $p$
arguments in our notation with the understanding that $p=p_c^{(n)}$.)
As was recently first pointed out by Hovi and Aharony~\cite{hovi97},
the $n$-cell iteration of Eq.~(\ref{eq:XRGc}), with the distribution
of $X_n$ fixed in the upper limit of the random sums, corresponds to a
stationary Watson-Galton branching
process~\cite{harris48,harris,athreya,feller}.  This classical discrete
branching process describes the growth of a population of individuals,
each of whom has a random number of children, $X_n$, sampled
independently from the same distribution.

The self-similar iteration of Eq.~(\ref{eq:XRGc}), 
\begin{equation}
X_{m^2} = \sum_{i=1}^{X_m} X_m^{(i)},
\end{equation}
starting from the initial condition $X_n$ (where $N = n^{2^j}$)
corresponds to a hierarchical sampling of the Watson-Galton process:
The distribution of grandchildren is that of the children's children,
the distribution of great, great grandchildren is that of the
grandchildren's grandchildren, etc. In this form, the SRG at the
critical fixed point clearly expresses {\it statistical
self-similarity at all length scales} because the coarse-grained mass
``on the cells'' (the random number of terms in the sum) has the same
distribution as the random mass contributions from each cell.

In terms of cumulant generating functions, the SRG at the critical
fixed point is simply a functional iteration,
\begin{equation}
\hat{f}_{mn} = \hat{f}_n \circ \hat{f}_m, \label{eq:critfRG}
\end{equation}
starting from the initial condition $\hat{f}_n(s)$.  From
Eq.~(\ref{eq:muRG}), this implies that the mean scales with system
size, $N = L^d = n^j$,  as a power law,
\begin{equation}
\mu_{n^j} = (\mu_n)^j = L^{D_f^{(n)}} = (n^{D_f^{(n)}/d})^j \label{eq:mun}
\end{equation}
where the $n$-cell estimate of the fractal dimension is given once again
by Eq.~(\ref{eq:Df}). 

(From Eq.~(\ref{eq:mun}), we see that $\mu_n>1$ is a necessary
condition for the expected value of the critical order parameter to
grow with system size, or equivalently, for the fractal dimension to
be positive. In the context of population growth, this is the case
where the expected population size grows exponentially with the number
of generations, e.g. a supercritical nuclear reaction. Conversely, the
case, $\mu_n <1$, models exponentially fast extinction of a
population, e.g. a subcritical nuclear reaction. As such, probability
theorists refer to these two types of branching processes as
``supercritical'' and ``subcritical'', respectively, and while the
marginal case $\mu_n=1$ is naturally termed
``critical''~\cite{feller,harris,athreya}.  Unfortunately, this
standard mathematical terminology is poorly suited here because the
SRG at the critical fixed point is actually a ``supercritical
branching process''. In fact, the non-stationary SRG branching process
is actually ``supercritical'' for any thermodynamic phase since the
expected value of the order-parameter always increases with system
size.)

\subsection{ The Fractal Central Limit Theorem }

Although the SRG corresponds to a new kind of non-stationary branching
process, various existing mathematical results for simple
Watson-Galton branching processes~\cite{harris48,harris,athreya} have
direct relevance for the critical fixed point~\cite{hovi97}.  For
example, the ``central region'' of the limiting distribution is at the
same scale as the mean.  In other words, the scaled random variable,
$Z_N = X_N/\mu_N$, has a stationary limiting distribution,
\begin{equation}
\lim_{j\rightarrow\infty} \mbox{Prob.}(X_{n^j} < z\; \mu_{n^j}) =
G^{(n)}(z)
\end{equation}
for some nontrivial function $G^{(n)}(z)$ (the $n$th SRG estimate of
the true limiting distribution). This result, which can be derived
from Eq.~(\ref{eq:critfRG}), explains why random fractals, such as
critical spanning clusters in percolation, have very large mass
fluctuations, comparable to the mean, in contrast to normal random
objects, such as supercritical spanning clusters, which have
``square-root fluctuations'' (at the scale of the square root of the
mean) as a consequence of the CLT.

\begin{figure}
\begin{center}
\mbox{ 
\psfig{file=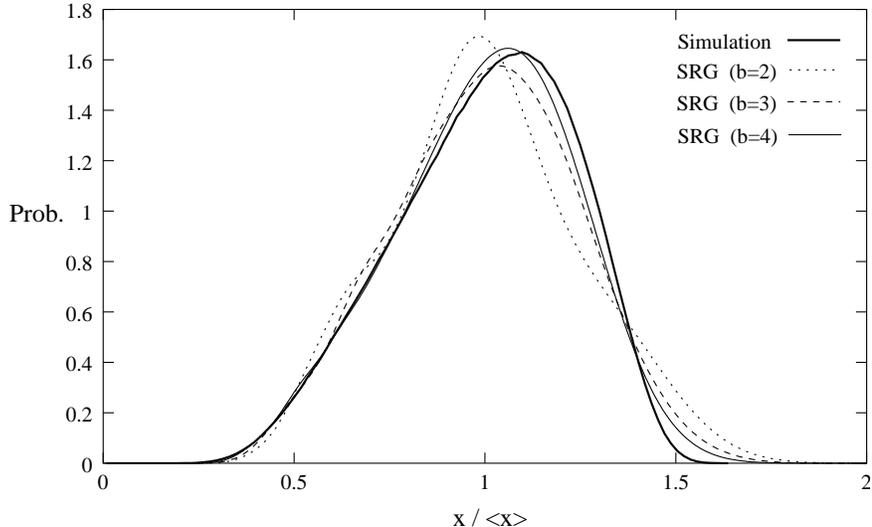,width=4.5in}
}
\begin{minipage}[h]{5.5in}
\caption{ \label{fig:compare-spanx} Limiting probability distribution
(scaled to have unit mean) for the total mass of clusters spanning a
square system in one direction (solid line, obtained from simulations
on a $300\times 300$ site lattice with $p=0.592746 \approx p_c$)
compared with $2\times 2$, $3 \times 3$, and $4\times 4$ SRG
approximations (dashed lines, obtained by numerically iterating the
exact recursions given in the main text).  }
\end{minipage}
\end{center}
\end{figure}

In Part III, it is shown that for percolation quantities, the SRG
critical fixed point, $G(z) = \lim_{n\rightarrow\infty} G^{(n)}(z)$,
is generally not Gaussian. Its shape depends sensitively on
macroscopic constraints, such as spanning or boundary conditions,
although it is presumably universal for different lattices in the same
embedding dimension. The fact that shape of the critical
order-parameter distribution depends on macroscopic constraints, for
arbitrarily large systems, is a striking signature of long-range
correlations, which has apparently not been previously noted in the
literature. This is also the first conclusion of the FCLT: {\it In the
central region, the limiting mass distribution of a random unifractal
is non-Gaussian and depends sensitively on macroscopic constraints.}

Preliminary numerical evidence indicates that satisfactory
approximations of critical distribution functions can be obtained
using remarkably small cells. These approximations are produced
by numerically iterating the recursions given above, either for the
probability density directly (which involves discrete convolutions) or
for one of the generating functions (which involves simple
multiplications followed by an inverse transform).  In
Fig.~\ref{fig:compare-spanx}, we see that the very simple $n = 2
\times 2$ cell scheme discussed above is surprisingly close to the
results of large-scale simulations for clusters spanning a square site
lattice in one direction. Even more striking is the good accuracy
achieved using only slightly larger cells, $n=3\times 3$ and $n =
4\times 4$. These approximations and others are analyzed in more
detail in Part III.

The SRG also predicts universal stretched-exponential decay for the
tails of the critical limiting distributions,
\begin{equation}
\log g(z) \sim \left\{ \begin{array}{ll} - A(z) z^{\hat{\delta}} & \mbox{ as
} z \rightarrow \infty \\ - B(z) z^{\hat{\eta}} & \mbox{ as } z \rightarrow 0
\end{array}\right.
\label{eq:gdecay}
\end{equation}
where $g(z) = G^\prime(z)$ is the limiting probability density
function in the limit of infinite cell size, and $A(z)$ and $B(z)$ are
periodic functions of $log z$, which are essentially constant in
practice~\cite{hovi97}. (Note that stretched-exponential decay of the
left tail as $z \rightarrow 0$ holds only if $x_n^- > 1$.) The tail
exponents, $\hat{\delta}$ and $\hat{\eta}$, can be expressed in terms
of other critical exponents and do not depend on macroscopic
constraints. If the mean scales like $\mu_n(p_c) \sim n^{\delta}$ and
the largest possible value scales like $x_n^+ \sim n^{\delta^+}$, then
the exponent for the right tail is given by,
\begin{equation}
\hat{\delta} = \frac{\delta^+}{\delta^+ -\delta}.
\label{eq:deltahat}
\end{equation}
Similarly, if the smallest possible value scales like $x_n^- \sim
n^{\delta^-}$, then the exponent for the left tail is given by,
\begin{equation}
\hat{\eta} = \frac{\delta^-}{\delta^- - \delta }.
\end{equation}
In the case of spanning clusters in $d$ dimensions, we have $\delta =
D_f/d$, $\delta^+ = 1$, and $\delta^- = 1/d$, in which case the tail
exponents are,
\begin{equation}
\hat{\delta} = \frac{d}{d-D_f}
\end{equation}
and
\begin{equation}
\hat{\eta} = \frac{1}{1 - D_f}. \label{eq:etahat}
\end{equation}
These conclusions follow directly from classical mathematical theorems
on Watson-Galton branching processes~\cite{harris,athreya,harris48},
as first pointed out by Hovi and Aharony~\cite{hovi97}. Note that the
co-dimension, $d-D_f$, governs the right tail, because rare, very
large clusters have the dimension of the system $d$, rather than the
normal fractal dimension, $D_f$. Similarly, the co-dimension, $D_f-1$,
governs in the left tail, because rare, very small clusters
(conditional on spanning) are roughly linear objects with dimension
one.

These results lead us to the second general conclusion of the FCLT:
{\it Outside the central region, the limiting distributions of random
unifractals have stretched exponential tails with universal exponents
given by Eqs.~(\ref{eq:deltahat})--(\ref{eq:etahat})}.  In ongoing
work~\cite{bb02}, the generality of the FCLT is explored by
comparisons with percolation, random graphs, and the Ising model, and
the subtle question of what happens above the upper
critical dimension is also addressed.

\begin{figure}
\begin{center}
\mbox{
\psfig{file=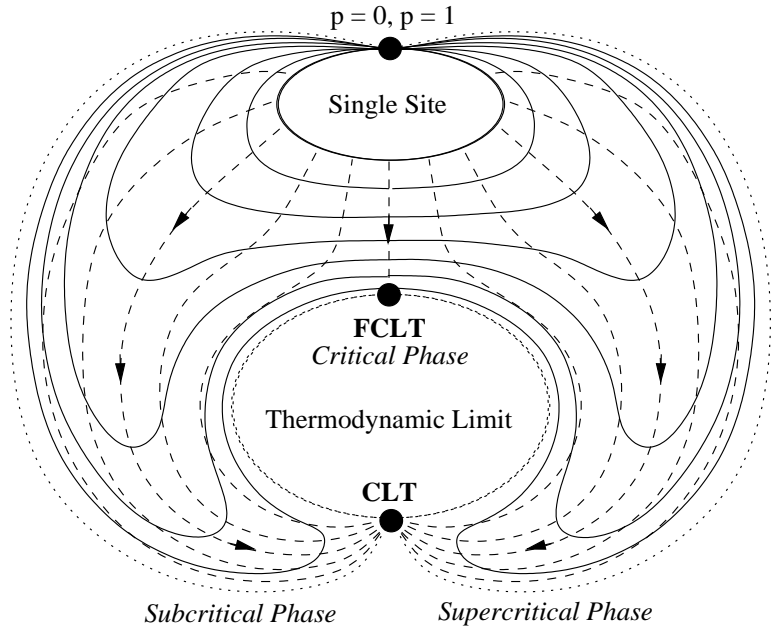,width=4in}
}
\begin{minipage}[h]{5.5in}
\caption{\label{fig:RGflow-span} Topology of the SRG flow for in the
space of scaled probability distributions of any structural quantity
conditional on spanning, such as the total cluster mass or the minimum
chemical distance connecting two opposite boundaries.  }
\end{minipage}
\end{center}
\end{figure}

\section{ Discussion }
\label{sec:disc}

Combining the results of this article, we arrive at a fruitful
intuitive picture of the SRG ``flow" (dashed lines) in the space of
probability distributions, as sketched in
Fig.~\ref{fig:RGflow-span}. As the system size is increased, the
physical manifold $0 \leq p \leq 1$ (solid lines) is advected from the
single-site manifold for $N=1$ toward the thermodynamic manifold for
$N = \infty$ (interior dotted line).  The SRG accurately describes the
universal flow near the two thermodynamic fixed points and the
crossover manifold connecting them, far away from the $p=0,1$ fixed
point representing discreteness and non-universality. For a fixed
value of $p \neq 0, p_c, 1$, at small sizes the system trajectory
passes near the (unstable) critical fixed point where the FCLT holds
and then crosses over toward the (stable) off-critical fixed point,
where the CLT holds, once the system size exceeds the correlation
length.



The significance of the SRG is both practical and fundamental.  In
real applications, critical phenomena such as percolation occur in a
finite systems, which are often small enough compared to the
correlation length that finite-size effects are  important. In
such cases, the SRG provides a simple analytical or numerical method
to approximate the mean, variance, and distribution of an order
parameter for arbitrary system sizes and values of the control
parameter, even when the correlation length and system size are
comparable. These approximations should be fairly accurate as long as
the correlation length and the system size are much larger than one.

Preliminary numerical results for percolation suggest that reasonable
accuracy can be achieved with remarkably small cell sizes ($b >
2$). Developing the best small-cell approximations for various
quantities in percolation and other statistical models will require
more extensive comparisons with simulation data (as in upcoming parts
of this series). Whenever reasonable small-cell schemes can be 
identified, however, they allow trivial calculations which may suffice to
replace direct simulations in many applications. From a practical
point of view, simple approximations for a broad range of conditions
can be much more useful than exact results for certain limiting cases
because models like percolation are always themselves only crude
approximations of real physical systems.

The SRG also provides basic insights into critical phenomena and
random fractals. The connection with branching processes gives a
simple ``population-growth'' picture of how crossover occurs in a
continuous phase transition.  The fixed points of the SRG also explain
the nature of fluctuations in different thermodynamic phases,
e.g. governed by the CLT and the FCLT.  These conclusions have been
reached here in the specific context of percolation, which is perhaps
the simplest case, but it should be possible to extend the SRG to
other critical phenomena, such as the Ising model, by postulating and
analyzing appropriate non-stationary branching processes expressing
statistical self-similarity.

\section*{ Acknowledgments }

The author gratefully acknowledges L. K. Bazant, J.-P. Bouchaud,
L. P. Kadanoff, P. Sen, and H. E. Stanley for useful discussions, and
D. Harmon for also providing the small-cell SRG results presented in
some of the figures.


\begin{thebibliography}{66}
\expandafter\ifx\csname natexlab\endcsname\relax\def\natexlab#1{#1}\fi
\expandafter\ifx\csname bibnamefont\endcsname\relax
  \def\bibnamefont#1{#1}\fi
\expandafter\ifx\csname bibfnamefont\endcsname\relax
  \def\bibfnamefont#1{#1}\fi
\expandafter\ifx\csname citenamefont\endcsname\relax
  \def\citenamefont#1{#1}\fi
\expandafter\ifx\csname url\endcsname\relax
  \def\url#1{\texttt{#1}}\fi
\expandafter\ifx\csname urlprefix\endcsname\relax\def\urlprefix{URL }\fi
\providecommand{\bibinfo}[2]{#2}
\providecommand{\eprint}[2][]{\url{#2}}

\bibitem{stanley}
\bibinfo{author}{\bibfnamefont{H.~E.} \bibnamefont{Stanley}},
  \emph{\bibinfo{title}{Introduction to Phase Transitions and Critical
  Phenomena}} (\bibinfo{publisher}{Oxford University Press},
  \bibinfo{year}{1971}).

\bibitem{ma}
\bibinfo{author}{\bibfnamefont{S.-K.} \bibnamefont{Ma}},
  \emph{\bibinfo{title}{Modern Theory of Critical Phenomena}}
  (\bibinfo{publisher}{Perseus Publishing, Cambridge, MA},
  \bibinfo{year}{1976}).

\bibitem{goldenfeld}
\bibinfo{author}{\bibfnamefont{N.}~\bibnamefont{Goldenfeld}},
  \emph{\bibinfo{title}{Lectures on Phase Transitions and the Renormalization
  Group}} (\bibinfo{publisher}{Perseus Books, Reading, MA},
  \bibinfo{year}{1992}).

\bibitem{kadanoff}
\bibinfo{author}{\bibfnamefont{L.~P.} \bibnamefont{Kadanoff}},
  \emph{\bibinfo{title}{Statics, Dyanmics and Renormalization}}
  (\bibinfo{publisher}{World Scientific, Singapore}, \bibinfo{year}{2000}).

\bibitem{bunde}
\bibinfo{author}{\bibfnamefont{A.}~\bibnamefont{Bunde}} \bibnamefont{and}
  \bibinfo{author}{\bibfnamefont{S.}~\bibnamefont{Havlin}}, in
  \emph{\bibinfo{booktitle}{Fractals and Disordered Systems}}, edited by
  \bibinfo{editor}{\bibfnamefont{A.}~\bibnamefont{Bunde}} \bibnamefont{and}
  \bibinfo{editor}{\bibfnamefont{S.}~\bibnamefont{Havlin}}
  (\bibinfo{publisher}{Springer}, \bibinfo{year}{1996}), pp.
  \bibinfo{pages}{58--113}, \bibinfo{edition}{2nd} ed.

\bibitem{hughes}
\bibinfo{author}{\bibfnamefont{B.}~\bibnamefont{Hughes}},
  \emph{\bibinfo{title}{Random Walks and Random Environments}},
  vol.~\bibinfo{volume}{II} (\bibinfo{publisher}{Clarendon Press, Oxford},
  \bibinfo{year}{1995}).

\bibitem{stauffer}
\bibinfo{author}{\bibfnamefont{D.}~\bibnamefont{Stauffer}} \bibnamefont{and}
  \bibinfo{author}{\bibfnamefont{A.}~\bibnamefont{Aharony}},
  \emph{\bibinfo{title}{Introduction to Percolation Theory}}
  (\bibinfo{publisher}{Taylor and Francis, London}, \bibinfo{year}{1994}),
  \bibinfo{edition}{2nd} ed.

\bibitem{havlin85}
\bibinfo{author}{\bibfnamefont{S.}~\bibnamefont{Havlin}},
  \bibinfo{author}{\bibfnamefont{B.}~\bibnamefont{Trus}},
  \bibinfo{author}{\bibfnamefont{G.~H.} \bibnamefont{Weiss}}, \bibnamefont{and}
  \bibinfo{author}{\bibfnamefont{D.}~\bibnamefont{Ben-Avraham}},
  \bibinfo{journal}{J. Phys. A.} \textbf{\bibinfo{volume}{18}}
 (\bibinfo{year}{1985})   \bibinfo{pages}{L247}.

\bibitem{havlin87a}
\bibinfo{author}{\bibfnamefont{S.}~\bibnamefont{Havlin}} \bibnamefont{and}
  \bibinfo{author}{\bibfnamefont{D.}~\bibnamefont{Ben-Avraham}},
  \bibinfo{journal}{Adv. Phys.} \textbf{\bibinfo{volume}{36}},
  \bibinfo{pages}{695} (\bibinfo{year}{1987}).

\bibitem{havlin87}
\bibinfo{author}{\bibfnamefont{S.}~\bibnamefont{Havlin}},
  \bibinfo{author}{\bibfnamefont{J.~E.} \bibnamefont{Keifer}},
  \bibinfo{author}{\bibfnamefont{F.}~\bibnamefont{Leyvraz}}, \bibnamefont{and}
  \bibinfo{author}{\bibfnamefont{G.~H.} \bibnamefont{Weiss}},
  \bibinfo{journal}{J. Stat. Phys.} \textbf{\bibinfo{volume}{47}},
  \bibinfo{pages}{173} (\bibinfo{year}{1987}).

\bibitem{neumann88}
\bibinfo{author}{\bibfnamefont{A.~U.} \bibnamefont{Neumann}} \bibnamefont{and}
  \bibinfo{author}{\bibfnamefont{S.}~\bibnamefont{Havlin}},
  \bibinfo{journal}{J. Stat. Phys.} \textbf{\bibinfo{volume}{52}},
  \bibinfo{pages}{203} (\bibinfo{year}{1988}).

\bibitem{hovi95}
\bibinfo{author}{\bibfnamefont{J.-P.} \bibnamefont{Hovi}} \bibnamefont{and}
  \bibinfo{author}{\bibfnamefont{A.}~\bibnamefont{Aharony}},
  \bibinfo{journal}{Fractals} \textbf{\bibinfo{volume}{3}},
  \bibinfo{pages}{4553} (\bibinfo{year}{1995}).

\bibitem{hovi97}
\bibinfo{author}{\bibfnamefont{J.-P.} \bibnamefont{Hovi}} \bibnamefont{and}
  \bibinfo{author}{\bibfnamefont{A.}~\bibnamefont{Aharony}},
  \bibinfo{journal}{Phys. Rev. E} \textbf{\bibinfo{volume}{56}},
  \bibinfo{pages}{172} (\bibinfo{year}{1997}).

\bibitem{dok98}
\bibinfo{author}{\bibfnamefont{N.~V.} \bibnamefont{Dokholyan}},
  \bibinfo{author}{\bibfnamefont{Y.}~\bibnamefont{Lee}},
  \bibinfo{author}{\bibfnamefont{S.}~\bibnamefont{Buldyrev}},
  \bibinfo{author}{\bibfnamefont{S.}~\bibnamefont{Havlin}},
  \bibinfo{author}{\bibfnamefont{P.~R.} \bibnamefont{King}}, \bibnamefont{and}
  \bibinfo{author}{\bibfnamefont{H.~E.} \bibnamefont{Stanley}},
  \bibinfo{journal}{J. Stat. Phys.} \textbf{\bibinfo{volume}{93}},
  \bibinfo{pages}{603} (\bibinfo{year}{1998}).

\bibitem{dok99}
\bibinfo{author}{\bibfnamefont{N.~V.} \bibnamefont{Dokholyan}},
  \bibinfo{author}{\bibfnamefont{S.}~\bibnamefont{Buldyrev}},
  \bibinfo{author}{\bibfnamefont{S.}~\bibnamefont{Havlin}},
  \bibinfo{author}{\bibfnamefont{P.~R.} \bibnamefont{King}},
  \bibinfo{author}{\bibfnamefont{Y.}~\bibnamefont{Lee}}, \bibnamefont{and}
  \bibinfo{author}{\bibfnamefont{H.~E.} \bibnamefont{Stanley}},
  \bibinfo{journal}{Physica A} \textbf{\bibinfo{volume}{266}},
  \bibinfo{pages}{55} (\bibinfo{year}{1999}).

\bibitem{andrade00}
\bibinfo{author}{\bibfnamefont{J.~S.} \bibnamefont{Andrade}},
  \bibinfo{author}{\bibfnamefont{S.~V.} \bibnamefont{Buldreyev}},
  \bibinfo{author}{\bibfnamefont{N.~V.} \bibnamefont{Dokholyan}},
  \bibinfo{author}{\bibfnamefont{S.}~\bibnamefont{Havlin}},
  \bibinfo{author}{\bibfnamefont{P.~R.} \bibnamefont{King}},
  \bibinfo{author}{\bibfnamefont{Y.}~\bibnamefont{Lee}},
  \bibinfo{author}{\bibfnamefont{G.}~\bibnamefont{Paul}}, \bibnamefont{and}
  \bibinfo{author}{\bibfnamefont{H.~E.} \bibnamefont{Stanley}},
  \bibinfo{journal}{Phys. Rev. E} \textbf{\bibinfo{volume}{62}},
  \bibinfo{pages}{8270} (\bibinfo{year}{2000}).

\bibitem{paul02}
\bibinfo{author}{\bibfnamefont{G.}~\bibnamefont{Paul}},
  \bibinfo{author}{\bibfnamefont{S.}~\bibnamefont{Havlin}}, \bibnamefont{and}
  \bibinfo{author}{\bibfnamefont{H.~E.} \bibnamefont{Stanley}},
  \bibinfo{note}{cond-mat/0203092}.

\bibitem{sen99}
\bibinfo{author}{\bibfnamefont{P.}~\bibnamefont{Sen}}, \bibinfo{journal}{Int.
  J. Mod. Phys. C} \textbf{\bibinfo{volume}{10}}, \bibinfo{pages}{747}
  (\bibinfo{year}{1999}).

\bibitem{duxbury87}
\bibinfo{author}{\bibfnamefont{P.~M.} \bibnamefont{Duxbury}} \bibnamefont{and}
  \bibinfo{author}{\bibfnamefont{P.~L.} \bibnamefont{Leath}},
  \bibinfo{journal}{J. Phys. A} \textbf{\bibinfo{volume}{20}},
  \bibinfo{pages}{L411} (\bibinfo{year}{1987}).

\bibitem{bazant00}
\bibinfo{author}{\bibfnamefont{M.~Z.} \bibnamefont{Bazant}},
  \bibinfo{journal}{Phys. Rev. E} \textbf{\bibinfo{volume}{62}},
  \bibinfo{pages}{1660} (\bibinfo{year}{2000}).

\bibitem{zeifman00}
\bibinfo{author}{\bibfnamefont{M.~I.} \bibnamefont{Zeifman}} \bibnamefont{and}
  \bibinfo{author}{\bibfnamefont{D.}~\bibnamefont{Ingman}},
  \bibinfo{journal}{J. Appl. Phys.} \textbf{\bibinfo{volume}{88}},
  \bibinfo{pages}{76} (\bibinfo{year}{2000}).

\bibitem{sen01}
\bibinfo{author}{\bibfnamefont{P.}~\bibnamefont{Sen}}, \bibinfo{journal}{J. Phys. A}, {\bf 34}, 8477 (\bibinfo{year}{2001}).

\bibitem{king99}
\bibinfo{author}{\bibfnamefont{P.~R.} \bibnamefont{King}},
  \bibinfo{author}{\bibfnamefont{J.~S.} \bibnamefont{Andrade}},
  \bibinfo{author}{\bibfnamefont{S.}~\bibnamefont{Buldyrev}},
  \bibinfo{author}{\bibfnamefont{N.~V.} \bibnamefont{Dokholyan}},
  \bibinfo{author}{\bibfnamefont{S.}~\bibnamefont{Havlin}},
  \bibinfo{author}{\bibfnamefont{Y.}~\bibnamefont{Lee}}, \bibnamefont{and}
  \bibinfo{author}{\bibfnamefont{H.~E.} \bibnamefont{Stanley}},
  \bibinfo{journal}{Physica A} \textbf{\bibinfo{volume}{266}}
  (\bibinfo{year}{1999}).

\bibitem{lee99}
\bibinfo{author}{\bibfnamefont{Y.}~\bibnamefont{Lee}},
  \bibinfo{author}{\bibfnamefont{J.~S.} \bibnamefont{Andrade}},
  \bibinfo{author}{\bibfnamefont{S.}~\bibnamefont{Buldyrev}},
  \bibinfo{author}{\bibfnamefont{N.~V.} \bibnamefont{Dokholyan}},
  \bibinfo{author}{\bibfnamefont{S.}~\bibnamefont{Havlin}},
  \bibinfo{author}{\bibfnamefont{P.~R.} \bibnamefont{King}},
  \bibinfo{author}{\bibfnamefont{G.}~\bibnamefont{Paul}}, \bibnamefont{and}
  \bibinfo{author}{\bibfnamefont{H.~E.} \bibnamefont{Stanley}},
  \bibinfo{journal}{Phys. Rev. Lett.}  (\bibinfo{year}{1999}).

\bibitem{torquato}
\bibinfo{author}{\bibfnamefont{S.}~\bibnamefont{Torquato}},
  \emph{\bibinfo{title}{Random Heterogeneous Materials}}
  (\bibinfo{publisher}{Springer, New York}, \bibinfo{year}{2000}).

\bibitem{newman80}
\bibinfo{author}{\bibfnamefont{C.~M.} \bibnamefont{Newman}},
  \bibinfo{journal}{Commun. Math. Phys.} \textbf{\bibinfo{volume}{74}},
  \bibinfo{pages}{119} (\bibinfo{year}{1980}).

\bibitem{newman81a}
\bibinfo{author}{\bibfnamefont{C.~M.} \bibnamefont{Newman}} \bibnamefont{and}
  \bibinfo{author}{\bibfnamefont{L.~S.} \bibnamefont{Schulman}},
  \bibinfo{journal}{J. Stat. Phys.} \textbf{\bibinfo{volume}{26}},
  \bibinfo{pages}{613} (\bibinfo{year}{1981}).

\bibitem{newman83}
\bibinfo{author}{\bibfnamefont{C.~M.} \bibnamefont{Newman}},
  \bibinfo{journal}{Commun. Math. Phys.} \textbf{\bibinfo{volume}{91}},
  \bibinfo{pages}{75} (\bibinfo{year}{1983}).

\bibitem{cox81}
\bibinfo{author}{\bibfnamefont{J.~T.} \bibnamefont{Cox}} \bibnamefont{and}
  \bibinfo{author}{\bibfnamefont{G.}~\bibnamefont{Grimmett}},
  \bibinfo{journal}{J. Stat. Phys.} \textbf{\bibinfo{volume}{25}},
  \bibinfo{pages}{237} (\bibinfo{year}{1981}).

\bibitem{cox84}
\bibinfo{author}{\bibfnamefont{J.~T.} \bibnamefont{Cox}} \bibnamefont{and}
  \bibinfo{author}{\bibfnamefont{G.}~\bibnamefont{Grimmett}},
  \bibinfo{journal}{Ann. Probab.} \textbf{\bibinfo{volume}{12}},
  \bibinfo{pages}{514} (\bibinfo{year}{1984}).

\bibitem{bb02}
\bibinfo{author}{\bibfnamefont{M.~Z.} \bibnamefont{Bazant}} \bibnamefont{and}
  \bibinfo{author}{\bibfnamefont{J.-P.} \bibnamefont{Bouchaud}},
  \bibinfo{note}{in preparation}.

\bibitem{feller}
\bibinfo{author}{\bibfnamefont{W.}~\bibnamefont{Feller}},
  \emph{\bibinfo{title}{An Introduction to Probability Theory and Its
  Applications}}, vol.~\bibinfo{volume}{II} (\bibinfo{publisher}{Wiley, New
  York}, \bibinfo{year}{1971}), \bibinfo{edition}{2nd} ed.

\bibitem{harris}
\bibinfo{author}{\bibfnamefont{T.~E.} \bibnamefont{Harris}},
  \emph{\bibinfo{title}{The Theory of Branching Processes}}
  (\bibinfo{publisher}{Springer-Verlag, Berlin}, \bibinfo{year}{1963}).

\bibitem{athreya}
\bibinfo{author}{\bibfnamefont{K.~B.} \bibnamefont{Athreya}} \bibnamefont{and}
  \bibinfo{author}{\bibfnamefont{P.~E.} \bibnamefont{Ney}},
  \emph{\bibinfo{title}{Branching Processes}}
  (\bibinfo{publisher}{Springer-Verlag, New York}, \bibinfo{year}{1972}).

\bibitem{stauffer80}
\bibinfo{author}{\bibfnamefont{D.}~\bibnamefont{Stauffer}},
  \bibinfo{journal}{Z. Phys. B} \textbf{\bibinfo{volume}{32}},
  \bibinfo{pages}{89} (\bibinfo{year}{1980}).

\bibitem{family80}
\bibinfo{author}{\bibfnamefont{F.}~\bibnamefont{Family}} \bibnamefont{and}
  \bibinfo{author}{\bibfnamefont{A.}~\bibnamefont{Coniglio}},
  \bibinfo{journal}{J. Phys. A} \textbf{\bibinfo{volume}{13}},
  \bibinfo{pages}{L403} (\bibinfo{year}{1980}).

\bibitem{margolina84}
\bibinfo{author}{\bibfnamefont{A.}~\bibnamefont{Margolina}} \bibnamefont{and}
  \bibinfo{author}{\bibfnamefont{H.~J.} \bibnamefont{Herrmann}},
  \bibinfo{journal}{Physics Letters} \textbf{\bibinfo{volume}{104A}},
  \bibinfo{pages}{295} (\bibinfo{year}{1984}).

\bibitem{barenblatt87}
\bibinfo{author}{\bibfnamefont{G.~I.} \bibnamefont{Barenblatt}},
  \emph{\bibinfo{title}{Dimensional Analysis}} (\bibinfo{publisher}{Gordon and
  Breach, New York}, \bibinfo{year}{1987}).

\bibitem{reynolds77}
\bibinfo{author}{\bibfnamefont{P.~J.} \bibnamefont{Reynolds}},
  \bibinfo{author}{\bibfnamefont{W.}~\bibnamefont{Klein}}, \bibnamefont{and}
  \bibinfo{author}{\bibfnamefont{H.~E.} \bibnamefont{Stanley}},
  \bibinfo{journal}{J. Phys. C} \textbf{\bibinfo{volume}{10}},
  \bibinfo{pages}{L167} (\bibinfo{year}{1977}).

\bibitem{reynolds78}
\bibinfo{author}{\bibfnamefont{P.~J.} \bibnamefont{Reynolds}},
  \bibinfo{author}{\bibfnamefont{H.~E.} \bibnamefont{Stanley}},
  \bibnamefont{and} \bibinfo{author}{\bibfnamefont{W.}~\bibnamefont{Klein}},
  \bibinfo{journal}{J. Phys. A} \textbf{\bibinfo{volume}{11}},
  \bibinfo{pages}{L199} (\bibinfo{year}{1978}).

\bibitem{reynolds80}
\bibinfo{author}{\bibfnamefont{P.~J.} \bibnamefont{Reynolds}},
  \bibinfo{author}{\bibfnamefont{H.~E.} \bibnamefont{Stanley}},
  \bibnamefont{and} \bibinfo{author}{\bibfnamefont{W.}~\bibnamefont{Klein}},
  \bibinfo{journal}{Phys. Rev. B} \textbf{\bibinfo{volume}{21}},
  \bibinfo{pages}{1223} (\bibinfo{year}{1980}).

\bibitem{young75}
\bibinfo{author}{\bibfnamefont{A.~P.} \bibnamefont{Young}} \bibnamefont{and}
  \bibinfo{author}{\bibfnamefont{R.~B.} \bibnamefont{Stinchcombe}},
  \bibinfo{journal}{J. Phys. C} \textbf{\bibinfo{volume}{8}},
  \bibinfo{pages}{L535} (\bibinfo{year}{1975}).

\bibitem{kirkpatrick77}
\bibinfo{author}{\bibfnamefont{S.}~\bibnamefont{Kirkpatrick}},
  \bibinfo{journal}{Phys. Rev. B} \textbf{\bibinfo{volume}{15}},
  \bibinfo{pages}{1533} (\bibinfo{year}{1977}).

\bibitem{yuge78a}
\bibinfo{author}{\bibfnamefont{Y.}~\bibnamefont{Yuge}} \bibnamefont{and}
  \bibinfo{author}{\bibfnamefont{C.}~\bibnamefont{Murase}},
  \bibinfo{journal}{J. Phys. A} \textbf{\bibinfo{volume}{11}},
  \bibinfo{pages}{L83} (\bibinfo{year}{1978}).

\bibitem{yuge78b}
\bibinfo{author}{\bibfnamefont{Y.}~\bibnamefont{Yuge}}, \bibinfo{journal}{Phys.
  Rev. B} \textbf{\bibinfo{volume}{18}}, \bibinfo{pages}{1514}
  (\bibinfo{year}{1978}).

\bibitem{ziff00}
\bibinfo{author}{\bibfnamefont{M.~E.~J.} \bibnamefont{Newman}}
  \bibnamefont{and} \bibinfo{author}{\bibfnamefont{R.~M.} \bibnamefont{Ziff}},
  \bibinfo{journal}{Phys. Rev. Lett.} \textbf{\bibinfo{volume}{85}},
  \bibinfo{pages}{4104} (\bibinfo{year}{2000}).

\bibitem{hovi96}
\bibinfo{author}{\bibfnamefont{J.-P.} \bibnamefont{Hovi}} \bibnamefont{and}
  \bibinfo{author}{\bibfnamefont{A.}~\bibnamefont{Aharony}},
  \bibinfo{journal}{Phys. Rev. E} \textbf{\bibinfo{volume}{53}},
  \bibinfo{pages}{235} (\bibinfo{year}{1996}).

\bibitem{langlands92}
\bibinfo{author}{\bibfnamefont{R.~P.} \bibnamefont{Langlands}},
  \bibinfo{author}{\bibfnamefont{C.}~\bibnamefont{Pichet}},
  \bibinfo{author}{\bibfnamefont{P.}~\bibnamefont{Pouliot}}, \bibnamefont{and}
  \bibinfo{author}{\bibfnamefont{Y.}~\bibnamefont{Saint-Aubin}},
  \bibinfo{journal}{J. Stat. Phys.} \textbf{\bibinfo{volume}{67}},
  \bibinfo{pages}{553} (\bibinfo{year}{1992}).

\bibitem{grassberger92}
\bibinfo{author}{\bibfnamefont{P.}~\bibnamefont{Grassberger}},
  \bibinfo{journal}{J. Phys. A} \textbf{\bibinfo{volume}{25}},
  \bibinfo{pages}{5475} (\bibinfo{year}{1992}).

\bibitem{ziff92}
\bibinfo{author}{\bibfnamefont{R.~M.} \bibnamefont{Ziff}},
  \bibinfo{journal}{Phys. Rev. Lett.} \textbf{\bibinfo{volume}{669}},
  \bibinfo{pages}{2670} (\bibinfo{year}{1992}).

\bibitem{langlands94}
\bibinfo{author}{\bibfnamefont{R.~P.} \bibnamefont{Langlands}},
  \bibinfo{author}{\bibfnamefont{P.}~\bibnamefont{Pouliot}}, \bibnamefont{and}
  \bibinfo{author}{\bibfnamefont{Y.}~\bibnamefont{Saint-Aubin}},
  \bibinfo{journal}{Ann. Math. Stat.} \textbf{\bibinfo{volume}{30}},
  \bibinfo{pages}{1} (\bibinfo{year}{1994}).

\bibitem{stauffer94}
\bibinfo{author}{\bibfnamefont{D.}~\bibnamefont{Stauffer}},
  \bibinfo{author}{\bibfnamefont{J.}~\bibnamefont{Adler}}, \bibnamefont{and}
  \bibinfo{author}{\bibfnamefont{A.}~\bibnamefont{Aharony}},
  \bibinfo{journal}{J. Phys. A} \textbf{\bibinfo{volume}{27}},
  \bibinfo{pages}{L475} (\bibinfo{year}{1994}).

\bibitem{grop94}
\bibinfo{author}{\bibfnamefont{U.}~\bibnamefont{Gropengiesser}}
  \bibnamefont{and} \bibinfo{author}{\bibfnamefont{D.}~\bibnamefont{Stauffer}},
  \bibinfo{journal}{Physica A} \textbf{\bibinfo{volume}{210}},
  \bibinfo{pages}{320} (\bibinfo{year}{1994}).

\bibitem{hu94}
\bibinfo{author}{\bibfnamefont{C.~K.} \bibnamefont{Hu}}, \bibinfo{journal}{J.
  Phys. A} \textbf{\bibinfo{volume}{27}}, \bibinfo{pages}{L813}
  (\bibinfo{year}{1994}).

\bibitem{hu95}
\bibinfo{author}{\bibfnamefont{C.-K.} \bibnamefont{Hu}} \bibnamefont{and}
  \bibinfo{author}{\bibfnamefont{J.-A.} \bibnamefont{Chen}},
  \bibinfo{journal}{J. Phys. A} \textbf{\bibinfo{volume}{28}},
  \bibinfo{pages}{L73} (\bibinfo{year}{1995}).

\bibitem{hovi94}
\bibinfo{author}{\bibfnamefont{A.}~\bibnamefont{Aharony}} \bibnamefont{and}
  \bibinfo{author}{\bibfnamefont{J.-P.} \bibnamefont{Hovi}},
  \bibinfo{journal}{Phys. Rev. Lett.} \textbf{\bibinfo{volume}{72}},
  \bibinfo{pages}{1941} (\bibinfo{year}{1994}).

\bibitem{lorenz98}
\bibinfo{author}{\bibfnamefont{C.~D.} \bibnamefont{Lorenz}} \bibnamefont{and}
  \bibinfo{author}{\bibfnamefont{R.~M.} \bibnamefont{Ziff}},
  \bibinfo{journal}{J. Phys. A} \textbf{\bibinfo{volume}{31}},
  \bibinfo{pages}{8147} (\bibinfo{year}{1998}).

\bibitem{ziff99}
\bibinfo{author}{\bibfnamefont{R.~M.} \bibnamefont{Ziff}},
  \bibinfo{author}{\bibfnamefont{C.~D.} \bibnamefont{Lorenz}},
  \bibnamefont{and} \bibinfo{author}{\bibfnamefont{P.}~\bibnamefont{Kleban}},
  \bibinfo{journal}{Physica A} \textbf{\bibinfo{volume}{266}},
  \bibinfo{pages}{17} (\bibinfo{year}{1999}).

\bibitem{ziff02} R. M. Ziff and M. E. J. Newman, preprint, cond-mat/0203496. 

\bibitem{cardy92}
\bibinfo{author}{\bibfnamefont{J.~L.} \bibnamefont{Cardy}},
  \bibinfo{journal}{J. Phys. A} \textbf{\bibinfo{volume}{25}},
  \bibinfo{pages}{L201} (\bibinfo{year}{1992}).

\bibitem{landau-binder}
\bibinfo{author}{\bibfnamefont{D.~P.} \bibnamefont{Landau}} \bibnamefont{and}
  \bibinfo{author}{\bibfnamefont{K.}~\bibnamefont{Binder}},
  \emph{\bibinfo{title}{A Guide to Monte Carlo Simulations in Statistical
  Physics}} (\bibinfo{publisher}{Cambridge University Press},
  \bibinfo{year}{2000}).

\bibitem{watson1874}
\bibinfo{author}{\bibfnamefont{H.~W.} \bibnamefont{Watson}} \bibnamefont{and}
  \bibinfo{author}{\bibfnamefont{F.}~\bibnamefont{Galton}},
  \bibinfo{journal}{J. Anthropol. Inst. Great Britain Ireland}
  \textbf{\bibinfo{volume}{4}}, \bibinfo{pages}{138} (\bibinfo{year}{1874}).

\bibitem{fisher30}
\bibinfo{author}{\bibfnamefont{R.~A.} \bibnamefont{Fisher}},
  \emph{\bibinfo{title}{The Genetical Theory of Natural Selection}}
  (\bibinfo{publisher}{Clarendon Press, Oxford}, \bibinfo{year}{1930}).

\bibitem{schroedinger}
\bibinfo{author}{\bibfnamefont{E.}~\bibnamefont{Schroedinger}},
  \emph{\bibinfo{title}{Probability Problems in Nuclear Chemistry}}, vol.
  \bibinfo{volume}{51A} (\bibinfo{publisher}{Proc. Roy. Irish Acad.},
  \bibinfo{year}{1945}).

\bibitem{newman81b}
\bibinfo{author}{\bibfnamefont{C.~M.} \bibnamefont{Newman}} \bibnamefont{and}
  \bibinfo{author}{\bibfnamefont{A.~L.} \bibnamefont{Wright}},
  \bibinfo{journal}{Ann. Probab.} \textbf{\bibinfo{volume}{9}},
  \bibinfo{pages}{671} (\bibinfo{year}{1981}).

\bibitem{grassberger00}
\bibinfo{author}{\bibfnamefont{P.}~\bibnamefont{Grassberger}} \bibnamefont{and}
  \bibinfo{author}{\bibfnamefont{W.}~\bibnamefont{Nadler}}
  (\bibinfo{year}{2000}), \bibinfo{note}{cond-mat/0010265}.

\bibitem{harris48}
\bibinfo{author}{\bibfnamefont{T.~E.} \bibnamefont{Harris}},
  \bibinfo{journal}{Ann. Math. Stat.} \textbf{\bibinfo{volume}{41}},
  \bibinfo{pages}{474} (\bibinfo{year}{1948}).

\end{thebibliography}

%
%
\nopagebreak

\end{document}